\documentclass[fleqn,usenatbib]{mnras}

\usepackage[T1]{fontenc}

\DeclareRobustCommand{\VAN}[3]{#2}
\let\VANthebibliography\thebibliography
\def\thebibliography{\DeclareRobustCommand{\VAN}[3]{##3}\VANthebibliography}

\usepackage{graphicx}	
\usepackage{amsmath}	
\usepackage{amssymb}	
\allowdisplaybreaks

\newcommand{\m}{\,{\rm m}}	%
\newcommand{\kms}{\,{\rm km\,s^{-1}}}
\newcommand{\s}{\,{\rm s}}
\newcommand{\muas}{\,\mu{\rm as}}%
\newcommand{\mas}{\,{\rm mas}}%
\newcommand{\yr}{\,{\rm yr}}
\newcommand{\kpc}{\,{\rm kpc}}	
\newcommand{\erf}{\,{\rm erf}}	
\newcommand{\erfi}{\,{\rm erfi}}	

\title[Testing QUMOND with GCs]{Testing QUMOND theory with Galactic globular clusters in a weak external field}

\author[A.\ Sollima et al.]{
A.\ Sollima$^{1}$\thanks{Deceased.}\thanks{We dedicate this paper to the memory of our dear friend and colleague Antonio Sollima, who passed away prematurely a few months ago. Antonio was a generous and witty person, and a creative and brilliant scientist, whom we miss deeply. Antonio conceived, developed and carried out most of this work, which we have had the honour to finalize.},
C.\ Nipoti$^{2}$\thanks{E-mail: carlo.nipoti@unibo.it},
F.\ Calura$^{1}$,
R.\ Pascale$^{1}$,
H.\ Baumgardt$^{3}$
\\
$^{1}$INAF Osservatorio di Astrofisica e Scienza dello spazio di Bologna, 
via Gobetti 93/3, 40129 Bologna, Italy\\
$^{2}$Dipartimento di Fisica e Astronomia "Augusto Righi", Università di Bologna, via Gobetti 93/2, 40129 Bologna, Italy\\
$^{3}$School of Mathematics and Physics, University of Queensland, St Lucia, QLD 4072, Australia
}

\date{Accepted 2023 July 18. Received 2023 July 14; in original form 2023 May 16}

\pubyear{2023}

\begin{document}
\label{firstpage}
\pagerange{\pageref{firstpage}--\pageref{lastpage}}
\maketitle

\begin{abstract}
We developed self-consistent dynamical models of stellar systems in the framework of quasi-linear modified Newtonian dynamics (QUMOND).
The models are constructed from the anisotropic distribution function of Gunn \& Griffin (1979), combined with the modified Poisson equation defining this gravitation theory and take into account the external field effect. 
We have used these models, and their Newtonian analogues, to fit the projected density and the velocity dispersion profiles of a sample of 18 Galactic globular clusters, using the most updated datasets of radial velocities and Gaia proper motions. We have thus obtained, for each cluster, estimates of the dynamical mass-to-light ratio ($M/L$) for each theory of gravity. The selected clusters 
have accurate proper motions and a well sampled mass function down to the very low mass regime. This allows us to constrain the degree of anisotropy and to provide, from comparison with stellar evolution isochrones, a dynamics-independent estimate of the minimum mass-to-light ratio $(M/L)_{min}$.  
Comparing the best-fitting dynamical $M/L$ with $(M/L)_{min}$, we find that for none of the analyzed clusters the two gravity theories are 
significantly incompatible with the observational data, although for one of them (NGC\,5024) the dynamical $M/L$ predicted by QUMOND lies at $2.8\sigma$ below $(M/L)_{min}$.
Though the proposed approach suffers from some limitations (in particular the lack of a treatment of
mass segregation), the obtained results suggest that the kinematics of globular 
clusters in a relatively weak external field can be a powerful tool to prove alternative theories of gravitation.
\end{abstract}




\section{Introduction}
\label{intro_sec}

One of the most astonishing astronomical discovery of the 20th century has been the tension 
between the estimate of the mass budget through luminous and dynamical tracers, 
suggesting the existence of a large amount of "dark matter".
Such an evidence was first noticed by \citet{zwicky1933,zwicky1937} as an anomaly 
in the velocity of galaxies inside the Coma cluster, 
and later confirmed at all scales, in the rotation of galactic disks 
\citep{babcock1939,kent1986,korsaga2019}, 
in the mass-to-light ratio of dwarf spheroidal galaxies \citep{faber1983,walker2009}, in 
the gravitational lensing of background objects by galaxy clusters 
\citep{wu1998,clowe2006}, in the temperature distribution of hot gas in galaxies 
and clusters \citep{mushotzky1991,mulchaey1993}, 
and in the pattern of anisotropies in the cosmic microwave background 
\citep{smoot1992,planck2016}.
According to the last estimate provided by the Planck satellite 
\citep{planck2020} $\sim$85\% of the mass budget of the Universe is constituted 
by non-baryonic matter. 
The freedom in the distribution of the dark mass in general allows one to solve the aforementioned 
tensions, so the dark matter paradigm is very hard to disprove. 
Nevertheless, despite the large effort in searching for dark matter particles, 
there has been no well-established detection of dark matter from a direct detection 
experiment \citep{carmona2016,xenon2020}.

Since the early '80s, \citet{milgrom1983} hypothesized an alternative 
explanation, postulating a modification of the standard 
Newtonian gravitation law at the regime of very low accelerations 
($<10^{-10}\m/\s^{2}$), known as modified Newtonian dynamics (MOND).
In particular, in spherical symmetry, the MOND gravitational field $-{\nabla\phi_{\rm M}}$ is related to the Newtonian gravitational field $-{\nabla\phi_{\rm N}}$ by
\begin{equation}
{\nabla\phi_{\rm M}}~\mu(\|{\nabla\phi_{\rm M}}\|/a_{0})={\nabla\phi_{\rm N}},
\label{g_eq}
\end{equation}
where $a_{0}$ is a  characteristic acceleration ($a_{0}\approx 1.2 \times 10^{-10}\m/\s^{2}$) and $\mu$ is a function that smoothly runs from $\mu(x)\sim x$ when $x\ll 1$ to $\mu(x)\sim 1$ 
when $x\gg 1$.
 
So, any stellar system behaves as Newtonian when $\|\nabla\phi_{\rm N}\|\gg a_{0}$, while 
its gravitational field deviates from Newtonian prediction as 
$\|\nabla\phi_{\rm M}\|\sim \sqrt{a_{0}\|{\nabla\phi_{\rm N}}\|}>\|{\nabla\phi_{\rm N}}\|$ 
when $\|\nabla\phi_{\rm N}\|\ll a_{0}$. As a consequence, objects crossing regions 
characterized by low acceleration move faster in MOND than in Newtonian gravity,  
which potentially can solve the tension between luminous and dynamical mass without the need of 
dark matter.
Such a simple and elegant modification, beside solving the dark matter issue, 
naturally reproduces the scaling relations of rotation- and pressure-supported 
galaxies like the Tully-Fisher \citep{tully1977} and Faber-Jackson 
\citep{faber1976} relations.

Following Milgrom's original idea, a few non-relativistic \citep[see][]{bekenstein1984,milgrom2010,milgrom2023} and relativistic \citep[see][]{bekenstein2004,famaey2012,skordis2021} MOND Lagrangian theories have been developed, such that the non-relativistic MOND gravitational field behaves essentially as in eq.\ (\ref{g_eq}).
In this work, we consider in particular the non-relativistic formulation of \citet{milgrom2010}, dubbed quasi-linear MOND (QUMOND), where the 
potential can be derived from the canonical Poisson equation adopting a "fake" 
density derivable from the actual density and the corresponding Newtonian potential.  
This theory thus involves solving only linear differential equations, with one 
non-linear, algebraic step. 
The QUMOND potential $\phi_{\rm M}$ obeys the equation
\begin{equation}
\nabla^{2} \phi_{\rm M}=\nabla \cdot \left[ \nu(y) \nabla\phi_{\rm N} \right] \label{mond_pois_eqa}
\end{equation}
or
\begin{equation}
 \nabla^{2} \phi_{\rm M}=4\pi G \rho_{f},
\label{mond_pois_eqb}
\end{equation}
where  $y=\|{\nabla \phi_{\rm N}}\|/a_{0}$,
and $\rho_{f}\equiv\nabla\cdot\left[ \nu(y) \nabla\phi_{\rm N} \right]/(4\pi G)$ is the fake density which is the source term of the canonical Poisson equation for $\phi_{\rm M}$.
The function $\nu$ is linked to the interpolating function so that 
$\mu(x)\nu(x\mu(x))=1$.

Over the years, MOND has been remarkably successful, resisting several 
attempts of falsification \citep[see][]{sanders2002,bekenstein2009}.
However, some features of observed systems and astrophysical phenomena are problematic for MOND, such as the dynamics and lensing of  clusters of galaxies \citep{the1988,clowe2006,natarajan2008}, 
the survival of the Fornax dwarf galaxy globular cluster 
system \citep{nipoti2008}, the internal and systemic dynamics of ultra faint 
dwarf galaxies \citep{safarazadeh2021}, the rotation curve of bulge dominated 
galaxies \citep{fraternali2011}, the X-ray isophotes of bright elliptical 
galaxies \citep{buote1994,angus2008}, the vertical kinematics of the Milky 
Way \citep{nipoti2007b,lisanti2019}, the resolved kinematics of the ultra-diffuse galaxy
AGC114905 \citep{mancera2022} and the phenomenon of galaxy 
merging \citep{nipoti2007}.

Another powerful class of objects useful to test this modified theory of 
gravitation is constituted by globular clusters \citep[GCs; ][]{baumgardt2005}.
They are almost spherical systems whose kinematics is determined by the balance 
between the gravitational force and the pressure due to the internal motions of their stars.
Although the majority of GCs have dense cores and therefore internal 
accelerations exceeding $a_{0}$ by orders of magnitudes, the gravitational 
acceleration quickly decreases with the distance  often 
reaching values below $a_{0}$ in their outskirts. 
So, the kinematics of an isolated GC with a sufficiently large radius is 
expected to be significantly different in Newtonian and MOND theories.

GCs are immersed in the Milky Way gravitational field whose strength 
is larger than $a_{0}$ at Galactocentric distances $R_{GC}<12~\kpc$ (including 
$\sim 71\%$ of the Galactic GC system).
Note that, for a satellite stellar system such as a GC, the gravitational field that appears in the argument of $\mu$ in eq.\ \ref{g_eq} is due to the contributions of both the satellite and the host system.  
So, in MOND, even a uniform external field affects the internal kinematics of a 
stellar system \citep[the so-called "external field effect"; ][]{bekenstein1984}.
However, the few GCs populating the outer halo of the Milky Way feel negligible 
external accelerations ($\|{\bf a_{ext}}\|\ll a_{0}$) and are extended enough to 
show significant differences in their velocity dispersion profiles according to 
the Newtonian dynamics and MOND \citep{baumgardt2005,sollima2010,ibata2011a,ibata2011b}, and therefore 
constitute an ideal tool to test these theories.

Two GCs have been analysed with this purpose till now: Palomar 14 and NGC 2419.
\citet{jordi2009} compared the projected velocity dispersion of Palomar 14 calculated 
with radial velocities of 17 member stars with a set of $N$-body 
simulations, reporting that the expected velocity dispersion in MOND is more 
than three times higher than the observed value, and concluded that this evidence 
challenges MOND. 
On the other hand, \citet{gentile2010} claimed that the confidence level 
achievable using the small sample of stars used by \citet{jordi2009} does not 
allow one to draw any significant conclusion on the validity of MOND. 
Finally, \citet{sollima2012} performed $N$-body simulations of Palomar 14 in both Newtonian gravity 
and MOND investigating the effect of different assumptions on the mass-to-light ratio $M/L$, binary fraction, 
anisotropy in the stellar velocity distribution, and cluster orbit. 
Comparing mock observations constructed from these simulations and the 
spectroscopic sample of \citet{jordi2009}, they concluded that both Newtonian and MOND 
models acceptably reproduce observations, with MOND models preferring low $M/L$.
They also found that even the weak external acceleration ($\|{\bf a_{ext}}\|\simeq 0.16 a_{0}$) 
felt by this GC produces significant effects on its kinematics.

Deeper analyses have been conducted on NGC2419. This GC is located at 
$\approx$96 kpc from the Galactic centre, thus feeling a negligible external field 
($\|{\bf a_{ext}}\|<0.1a_{0}$). Moreover, it is massive enough 
\citep[$M \simeq9.71\times 10^{5} M_{\odot}$; ][]{baumgardt2018} to contain hundreds of bright stars with accurate radial velocities.
\citet{ibata2011a,ibata2011b} used a large sample of $\sim160$ radial velocities and compared their velocity 
distribution with a set of dynamical models including the effect of anisotropy and binary fraction.
They found that Newtonian models fit observations better than MOND providing likelihood larger by factor $10^{5}$.
These results on NGC2419 have been however criticized by \citet{sanders2012a,sanders2012b} who argued that polytropic MOND models provide 
a reasonable fit to the data and claimed that likelihood-based analyses {\it a)} are 
dependent on the choice of the model stellar phase-space distribution function and {\it b)} can be 
used in a comparative test between different classes of models, but cannot rule 
out a model which adequately reproduces observations. 

A sound test to modified gravity would require the use of {\it i)} a simple and 
robust observational quantity which is as much as possible independent on the underlying distribution function, and {\it ii)} models flexible enough to reproduce those dynamical properties of the system (e.g.\ anisotropy) whose 
degrees of freedom are all well constrained.

The shape and amplitude of the velocity 
dispersion profile (the main kinematic quantity used as observational 
constraint) in any dynamical model depend on the adopted $M/L$ and on the 
degree of anisotropy. So, for a given pair of density and velocity dispersion 
profile, and once the anisotropy profile is fixed from the distribution of transverse motions, 
different gravitation theories will predict different dynamical $M/L$.

The cluster $M/L$ is therefore an excellent tool to test gravitation theories. 
Indeed, as discussed above, for a given mass, MOND models predict velocity dispersions systematically higher than Newtonian ones.
This is a property set by construction (linked to the increased acceleration of 
MOND below $a_{0}$) and it is independent of the adopted distribution function.
So, for a given velocity dispersion, MOND models require lower masses (and therefore $M/L$) than Newtonian ones.

The $M/L$ of a GC can be derived also with a method that is completely 
independent of dynamics, i.e.\ by summing the masses of individual stars detectable in deep photometric studies. 
This last task requires, beside a deep and complete photometry, the knowledge of the cluster mass function (MF) down to the faintest (lowest mass) stars and a stellar isochrone with suitable age and metallicity.
Comparing the dynamical $M/L$ with such dynamics-independent $M/L$ is thus a powerful method to test gravitational theories on the scale of GCs.

Unfortunately, even the deepest photometric studies performed with the 
Hubble Space Telescope on a large number of GCs \citep{sarajedini2007} are limited to 
the inner halo (at $R_{GC}<30\kpc$). Similarly, the exquisite accuracy of proper motions 
provided by Gaia \citep[$\sigma_{\mu}=25~\muas/\yr$ at $V\sim 16$; ][]{gaia2018a} translate into several km/s beyond $R_{GC}\sim 25\kpc$.
For GCs in this distance range the acceleration exerted by the Milky Way gravitational field can be several times $a_{0}$ 
and the external field effect cannot be neglected. 

Dealing with the external field effect is technically simpler in QUMOND than, for instance, in the \citet{bekenstein1984} formulation of MOND (see, e.g., \citealt{lughausen2015} and \citealt{chae2022}), which makes QUMOND the natural choice if one wants to test MOND also with the GCs of the inner halo. These GCs are close enough to have well sampled MF down to the hydrogen burning 
limit \citep{paust2010,sollima2017b,ebrahimi2020,baumgardt2023} and proper motions with accuracies comparable with those of radial velocities for hundreds of stars \citep{gaia2021}. 

In this paper we present self-consistent dynamical models in the QUMOND theory that are analogues of those of \citet{gunn1979} in 
Newtonian gravity. We then use these models to derive the dynamical mass-to-light ratio in the V band ($M/L_V$)
by best fitting the line-of-sight and transverse velocity distributions provided by the most updated compilation of radial velocities \citep{baumgardt2018} 
and Gaia proper motions for a sample of 18 Galactic GCs located between 2.5 and 
18.5 kpc from the Galactic centre, in a regime of relatively weak ($0.3<a_{ext}/a_{0}<4.9$) 
external acceleration. 
The comparison with the $M/L_V$ derived independently 
using the observed MF, age, metallicity and theoretical isochrones is used to test 
both Newtonian and QUMOND theories.

In Sect. \ref{mod_sec} the models are presented. In Sect. \ref{obs_sec} we describe the selected sample of GCs and the dataset used in this analysis.
Sect. \ref{tech_sec} is devoted to the description of the algorithm used to 
derive the $M/L_{V}$ from dynamics and its lower limit set from stellar models, together with their corresponding uncertainties. 
The results are presented in Sect. \ref{res_sec} and discussed in Sect. \ref{concl_sec}.

\section{Models}
\label{mod_sec}

\subsection{Model description}
\label{mod_descr_sec}

For both Newtonian and QUMOND models we adopted the distribution function defined by 
\citet[][based on \citealt{michie1963} and \citealt{king1966}]{gunn1979} 

\begin{equation}
f(E,L)=\exp\left(-\frac{L^{2}}{2 \sigma_{K}^{2} r_{a}^{2}}\right)\left[\exp\left(-\frac{E}{\sigma_{K}^{2}}\right)-1\right],
\label{df_eq}
\end{equation}
which can be written as
\begin{equation}
f(r,v_{r},v_{t})=\exp\left(-\frac{v_{t}^{2} r^{2}}{2 \sigma_{K}^{2} r_{a}^{2}}\right)\left[\exp\left(-\frac{v_{r}^{2}+v_{t}^{2}}{2\sigma_{K}^{2}}-\frac{\phi(r)}{\sigma_{K}^{2}}\right)-1\right],
\label{df_eq_sph_coor}
\end{equation}
where $E$ and $L$ are the energy and angular momentum per unit mass, $r$ is the distance from the cluster centre, $\phi$ is the gravitational potential, 
$r_{a}$ is the characteristic radius beyond which orbits become significantly 
radially biased, $\sigma_{K}^{2}$ is an energy normalization, and $v_{r}$ and $v_{t}$ are the radial and tangential components of the velocity, respectively.
The above distribution function represents the phase-space density and can be 
integrated over the velocity domain to obtain, as functions of radius, the density 
\begin{equation}
\rho=4\pi\int_{0}^{\sqrt{-2\phi}}dv_{r}\int_{0}^{\sqrt{-2\phi-v_{r}^{2}}}dv_{t} v_{t} f(r,v_{r},v_{t}),\\
\label{dv3d}
\end{equation}
and the radial ($\sigma_{r}$) and tangential ($\sigma_{r}$) velocity dispersions, which are given by
\begin{equation}
\sigma_{r}^{2}=\frac{4\pi}{\rho}\int_{0}^{\sqrt{-2\phi}}dv_{r}v_{r}^{2}\int_{0}^{\sqrt{-2\phi-v_{r}^{2}}} dv_{t} v_{t} f(r,v_{r},v_{t}),
\label{dv3d_sigr}
\end{equation}
and
\begin{equation}
\sigma_{t}^{2} =\frac{4\pi}{\rho}\int_{0}^{\sqrt{-2\phi}}dv_{r}\int_{0}^{\sqrt{-2\phi-v_{r}^{2}}}dv_{t}  v_{t}^{3} f(r,v_{r},v_{t}),
\label{dv3d_sigt}
\end{equation}
respectively.
The differential equation linking the potential derivatives to the density is the
canonical Poisson equation 
\begin{equation}
\nabla^{2}\phi_{\rm N}=4\pi G \rho
\label{pois_eq}
\end{equation}
in the the Newtonian case, and eq.\ (\ref{mond_pois_eqb}) in the QUMOND case. In all our MOND models we adopt the so-called "simple" interpolating function $\mu(x)=x/(1+x)$ \citep{famaey2005}, whose corresponding $\nu$ function is
\begin{equation}
\nu(y)=1+\frac{2}{y+\sqrt{y^{2}+4y}}.
\label{nu_eq}
\end{equation}

In the Newtonian case eq.\ \ref{pois_eq}, coupled with eq. \ref{dv3d}, can be solved in a straightforward way starting from a boundary condition at the centre for the
potential $\phi=\phi_{0}$ and integrating eq.\ \ref{dv3d} outwards out to where the potential and the density vanish.
In the QUMOND case the situation is more complex because eq.\ \ref{dv3d} provides, 
for a given potential $\phi_{\rm M}$, the actual density $\rho$, 
while eq.\ \ref{mond_pois_eqb} requires the fake density $\rho_{f}$.
Note however that the relation between $\rho$ and $\rho_{f}$ can be derived by combining 
eq.s \ref{mond_pois_eqa} and \ref{mond_pois_eqb}:
\begin{equation}
\rho_{f}=\nu\rho+\frac{\nu'(\nabla\|\nabla\phi_{\rm N}\|)\cdot\nabla\phi_{\rm N}}{4\pi G a_{0}}.
\label{rhof_eq}
\end{equation}
So, the Newtonian field $-\nabla\phi_{\rm N}$ becomes the 
only quantity necessary to determine at each radial step $\rho_{f}$ and close 
the system of equations \ref{dv3d}, \ref{mond_pois_eqb} and \ref{rhof_eq} for a given 
boundary condition for the potential at the centre.
The natural choice is to adopt $\nabla\phi_{\rm N}=0$ at $r=0$ and then derive
the radial profile of $\nabla\phi_{\rm N}$ from eq.\ \ref{pois_eq} \citep[see][]{king1966,gunn1979}.

For a cluster immersed in an external field, the argument of the function $\nu$ is the magnitude of the total (internal plus external) gravitational field normalized to $a_{0}$.
Because of the vectorial nature of the acceleration and of the different symmetry of the 
internal and external acceleration field, the magnitude of the total acceleration varies 
with the angle with respect to the direction of the external acceleration.
This breaks the spherical symmetry of the system whose density/potential contours 
will be elongated.
This introduces an inconsistency with the distribution function adopted in eq. \ref{df_eq}.
Indeed, while the energy remains an integral of motion regardless of the system geometry, 
the angular momentum magnitude $L$ is not conserved in a non-spherical system.
However, in slightly flattened potentials, say with axis ratios $\gtrsim 0.9$, $L$ is conserved within a few percent \citep[][sect.\ 3.2.2, pag. 163]{binney2008}.
As we will see below (Sect.\ \ref{mod_prop_sec}), our models are in fact slightly flattened, so we neglect this issue.

The outer boundary condition for the MOND and Newtonian gravitational fields are, respectively, 
\begin{equation*}
\lim_{r\to\infty} \nabla\phi^{\rm M}=-{\bf a_{ext}^{\rm M}},
\label{glim_eq}
\end{equation*}
where ${\bf a_{ext}^{\rm M}}$ is the MOND external field,
and 
\begin{equation*}
\lim_{r\to\infty} \nabla\phi_{\rm N}=-{\bf a_{ext}^{\rm N}},
\label{glim_eq}
\end{equation*}
where ${\bf a_{ext}^{\rm N}}$ is the Newtonian external field.

It is convenient to define the internal gravitational potentials $\psi_{\rm M}$ and $\psi_{\rm N}$, such that 
\begin{equation*}
\nabla\psi_{\rm M}=\nabla\phi_{\rm M}+{{\bf a}_{ext}^{\rm M}}
\label{psi_eq}
\end{equation*}
and
\begin{equation*}
\nabla\psi_{\rm N}=\nabla\phi_{\rm N}+{{\bf a}_{ext}^{\rm N}}.
\label{psi_eq}
\end{equation*}
In order to account for the external field effect, in all the equations of this section we must replace  
$\nabla\phi_{\rm M}$ with $\nabla\psi_{\rm M}-{{\bf a}^{\rm M}_{ext}}$ and $\nabla\phi_{\rm N}$ with $\nabla\psi_{\rm N}-{{\bf a}^{\rm N}_{ext}}$. Remarkably, as pointed out by \citet{milgrom2010}, ${\bf a_{ext}^{\rm M}}$ drops from the equation for the internal potential $\psi_{\rm M}$, which thus depends on 
${{\bf a}_{ext}^{\rm N}}$, but not on ${{\bf a}_{ext}^{\rm M}}$. In practice, the QUMOND internal field $\psi_{\rm M}$ can be obtained by solving, with boundary condition $\nabla\psi_{\rm M}\to 0$ at infinity,  the equation 
\begin{equation}
    \nabla^{2} \psi_{\rm M}=4\pi G \rho_{f}
    \label{eq:psimond}
\end{equation}
where
\begin{equation}
\rho_{f}=\nu\rho+\frac{\nu'(\nabla\|\nabla\psi_{\rm N}-{\bf a_{ext}^{\rm N}}\|)\cdot(\nabla\psi_{\rm N}-{\bf a_{ext}^{\rm N}})}{4\pi G a_{0}}
\label{rhof_eq}
\end{equation}
\citep[see][]{milgrom2010,chae2022}.
As usual, $\psi_{\rm N}$ can be obtained by solving 
\begin{equation}
    \nabla^{2} \psi_{\rm N}=4\pi G \rho,
\end{equation}
with standard boundary conditions.

The internal potential $\psi_{\rm M}$ is not spherically symmetric, but will maintain a symmetry with respect to the direction of the external field.
So, it is possible to express all the involved quantities (density, potential, velocity dispersions, 
etc.) as functions of the spherical polar coordinates $r$ and $\theta$, where $0<\theta<\pi$ is the angle formed with the positive $z$ axis, which is taken to have the same direction and orientation of the external field, while there is no dependence on the azimuthal coordinate $\phi$. 
We write the Newtonian and MOND potentials and densities as combinations of $N$ Legendre polynomials
\begin{eqnarray}
\psi&=&\sum_{k=0}^{N} u_{k}(r) P_{k}(\theta),\nonumber\\
\rho&=&\sum_{k=0}^{N} g_{k}(r) P_{k}(\theta),
\label{defphi_eq}
\end{eqnarray} 
where the functions $u_{k}$ and $g_{k}$ can be found by applying Laplace's equation 
and the variation of constant formula \citep[see ][]{prendergast1970,wilson1975}, so that
\begin{equation}
u_{0}=\psi_{0}+4\pi G \left(\int_{0}^{r} r g_{0}~dr-\frac{1}{r}\int_{0}^{r} r^{2} g_{0}~dr\right),
\label{uk_eq}
\end{equation}
and
\begin{equation}
u_{k}=-\frac{4\pi G}{2k+1} \left(r^{k}\int_{r}^{\infty} r^{1-k} g_{k}~dr+r^{-1-k}\int_{0}^{r} r^{k+2} g_{k}~dr\right),
\end{equation}
where
\begin{equation}
g_{k}=\frac{2k+1}{2} \int_{0}^{\pi} \rho~P_{k}~\sin\theta~d\theta
\label{gk_eq}
\end{equation}
and $\psi_0$ is the central potential. 
The same coefficients for QUMOND models can be calculated by replacing $\rho$ with $\rho_{f}$ in eq.s \ref{gk_eq}. 

The gradient of the internal Newtonian potential is therefore
\begin{equation*}
\begin{aligned}
\nabla\psi_{\rm N}=&\left(\sum_{k=0}^{N} \frac{d u_{k}}{dr}-\|{\bf a_{ext}^{\rm N}}\| \cos\theta\right) {\bf \hat{e}_{r}}\nonumber\\
&+\left(\sum_{k=0}^{N} \frac{u_{k}}{r}\frac{d P_{k}}{d\theta}-\|{\bf a_{ext}^{\rm N}}\| \sin\theta\right) {\bf \hat{e}_{\theta}}.\nonumber
\end{aligned}
\label{defpsi_eq}
\end{equation*} 

The model is computed iteratively, starting from $N=0$ (for which the model is spherical, 
$g_{0}=\rho$ and eq.\ \ref{uk_eq} is simply the canonical Poisson equation in its integral form).
The density profile of the $N=0$ model is then used to compute the fake density in the $(r,\theta)$ plane (from eq. \ref{rhof_eq}), 
the QUMOND potential (eq. \ref{mond_pois_eqb}), 
the actual density and velocity dispersions (eq. \ref{dv3d}), 
the high-order asymmetric components $g_{k}$ and $u_{k}$ (eq.s \ref{gk_eq}), 
and a new model is computed.
We found that $N=5$ provides reasonably stable models with only negligible differences with respect to higher-order models.
Note that a symmetric Newtonian potential $\psi_{\rm N}$, because of the presence of the external field, produces an asymmetric fake density
profile along the direction of the external field. So, the density distributions of 
subsequent iterations are shifted along this direction.
The updated density distribution is then shifted to bring the system centre to the origin of the axes and 
used as input to construct the models of the next iteration.
The density profiles of different steps are then compared and a new iteration is started if 
the average variation exceeds 0.1\% of the central density.

The model is then projected in the plane of the sky and the observational quantities (projected density $\Sigma$ and 
velocity dispersions along the line of sight $\sigma_{\rm LOS}$, projected radial $\sigma_{R}$ and tangential $\sigma_{T}$ directions)  are calculated.

In practice, it is convenient to express all quantities as dimensionless by normalizing the 
densities to the central value $\rho_{0}$ of $\rho$, the radii and the potential to characteristic values 
($r_{c}$ and $\sigma_{K}^{2}$) and the external acceleration to $a_{0}$:
\begin{align}
\tilde{\rho}&=\rho/\rho_{0},&\tilde{\rho}_{f}&=\rho_{f}/\rho_{0},\nonumber\\
\tilde{r}&=r/r_{c},&W&=-\psi/\sigma_{K}^{2},\nonumber\\
\tilde{a}&=|{\bf a_{ext}^{\rm N}}|/a_{0},&\tilde{r}_{a}&=r_{a}/r_{c}.
\label{norm_eq}
\end{align}

The shape of each QUMOND model is completely defined by five parameters: the central dimensionless potential $W_{0}$, the parameter $\xi=\sigma_{K}^{2}/(a_{0}r_{c})$, 
the strength of the external acceleration $\tilde{a}$ and the anisotropy radius $\tilde{r}_{a}$, 
and the cluster mass $M$, which determines both $\sigma_{K}^{2}$ and $r_{c}$ through the relations
\begin{equation}
r_{c}=\sqrt{\frac{4\pi G M}{9 I \xi a_{0}}},
\end{equation}
\begin{equation}
\sigma^2_{K}=\sqrt{\frac{4\pi G M \xi a_{0}}{9 I}},
\label{sigk_eq}
\end{equation}
and 
\begin{equation}
\rho_{0}=\frac{9\sigma_K^2}{4\pi G r_c^2},
\end{equation}
where
\begin{equation*}
I=2\pi\int_{0}^{\pi}\int_{0}^{\infty}\tilde{r}^2 \sin\theta \tilde{\rho} d\tilde{r} d\theta
\label{i_eq}
\end{equation*} 
\citep[see also ][]{sollima2010}.

In appendix we report more details about the
computation of the models (Appendix \ref{sec:app_mod}) and of the external acceleration 
(Appendix \ref{sec:app_ext}).

\begin{figure*}
\includegraphics[trim={0 5.5cm 0cm 0cm},clip,width=0.8\textwidth]{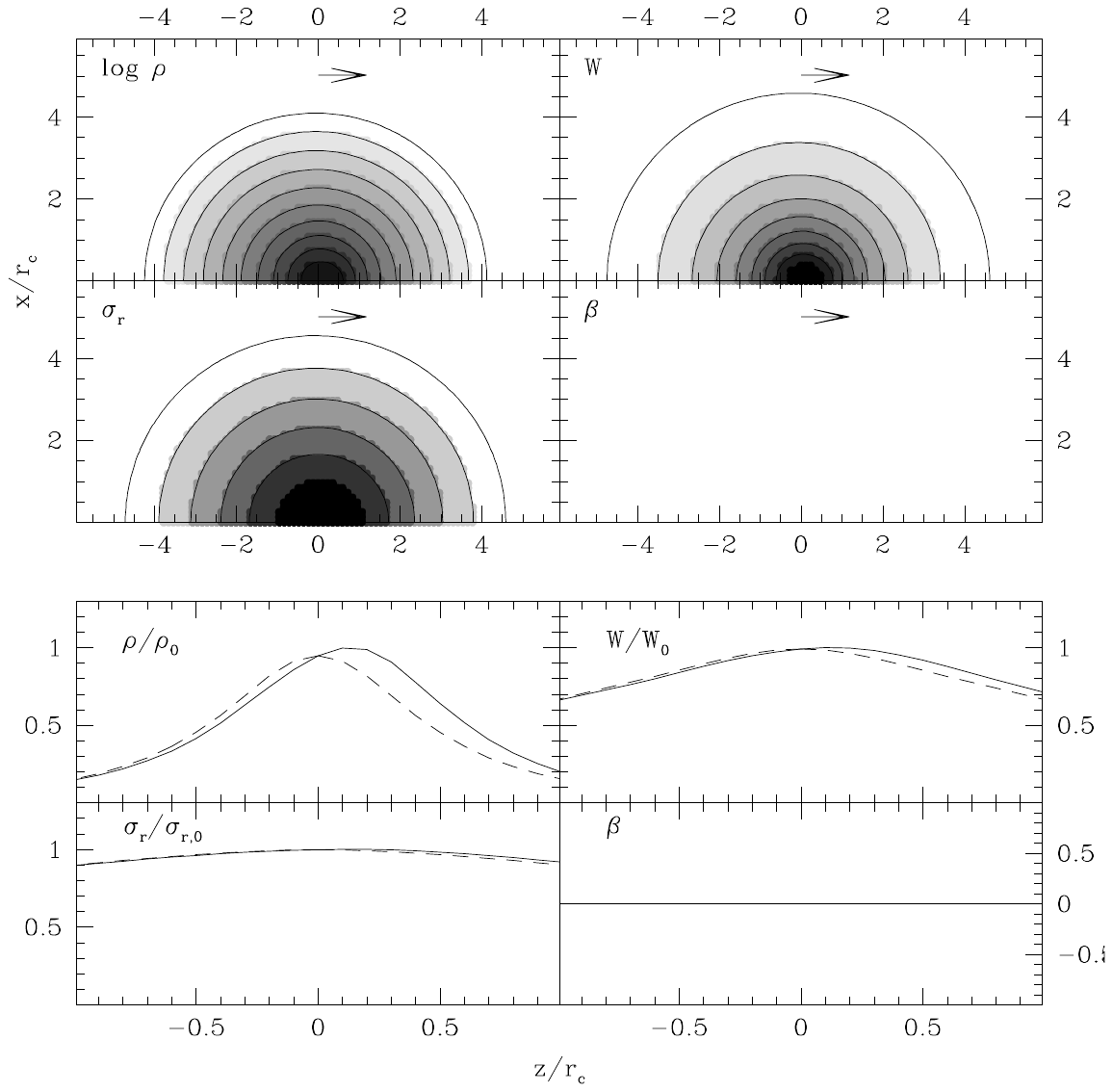}
\caption{Properties of the reference isotropic model (from top to bottom). First row: maps of the logarithmic density (left panel) and
dimensionless potential (right panel). Second row: radial velocity dispersion (left panel) and 
anisotropy parameter (right panel). 
Darker contours indicate larger values in steps of 10\% of the central value.
The direction of the external acceleration is shown by an arrow.
Third row: central density (left panel) and potential (right panel) profiles, along (solid lines) and orthogonal to (dashed lines) 
the direction of the external field.
Fourth row: radial velocity dispersion (left panel) and anisotropy parameter (right panel; here $\beta=0$ everywhere) profiles, along (solid lines) and orthogonal to (dashed lines) the direction of the external field.}
\label{mod1_fig}
\end{figure*}

\begin{figure*}
\includegraphics[trim={0 5.5cm 0cm 0cm},clip,width=\textwidth]{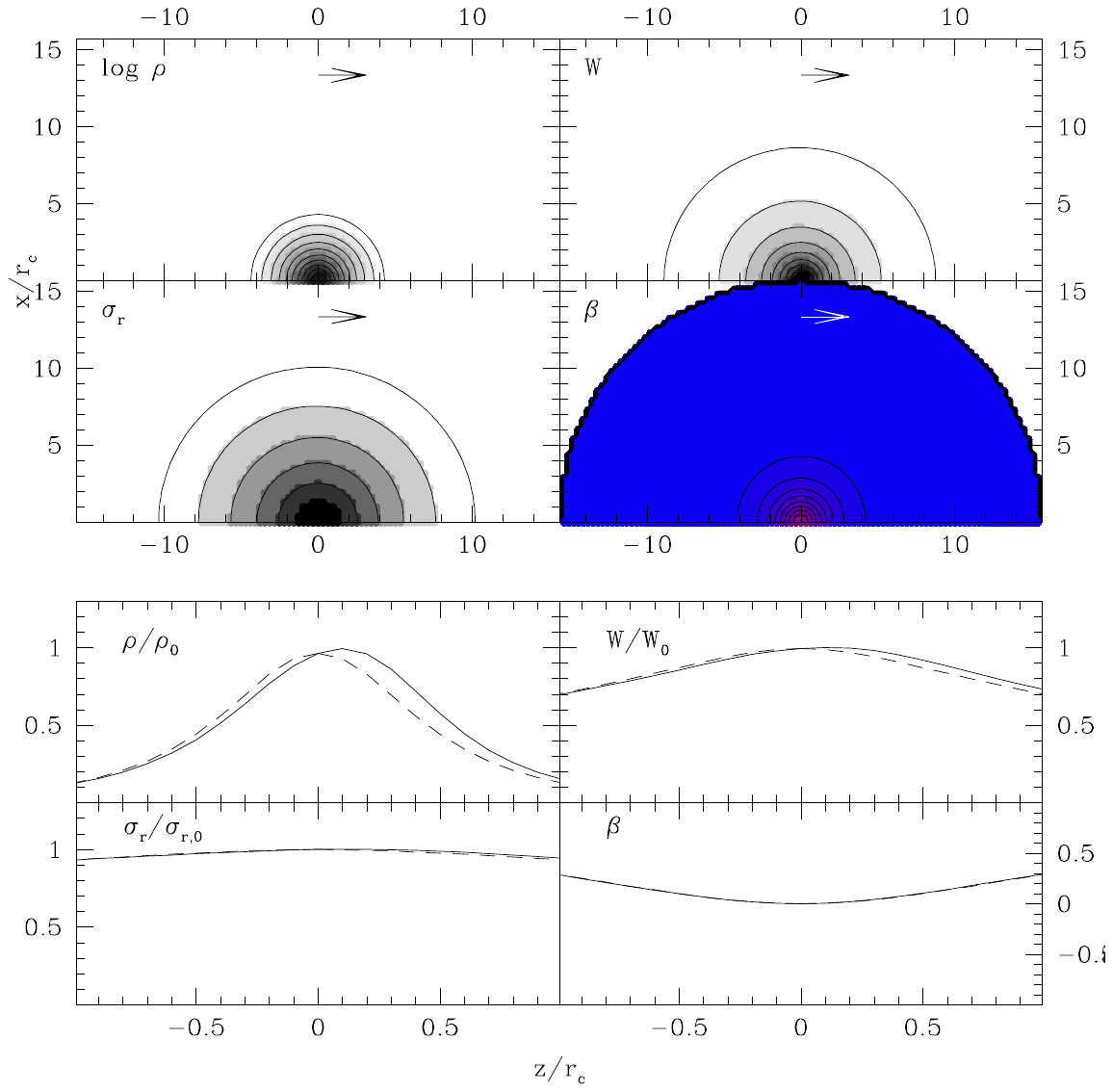}
\caption{Same as Fig. \ref{mod1_fig}, but for the anisotropic reference model.
In the map of the anisotropy parameter, colors range from red (more isotropic) to blue (more radially anisotropic).
}
\label{mod1a_fig}
\end{figure*}
 
\subsection{Model properties}
\label{mod_prop_sec}

\begin{figure}
\includegraphics[trim={0 5.5cm 0cm 0cm},clip,width=\columnwidth]{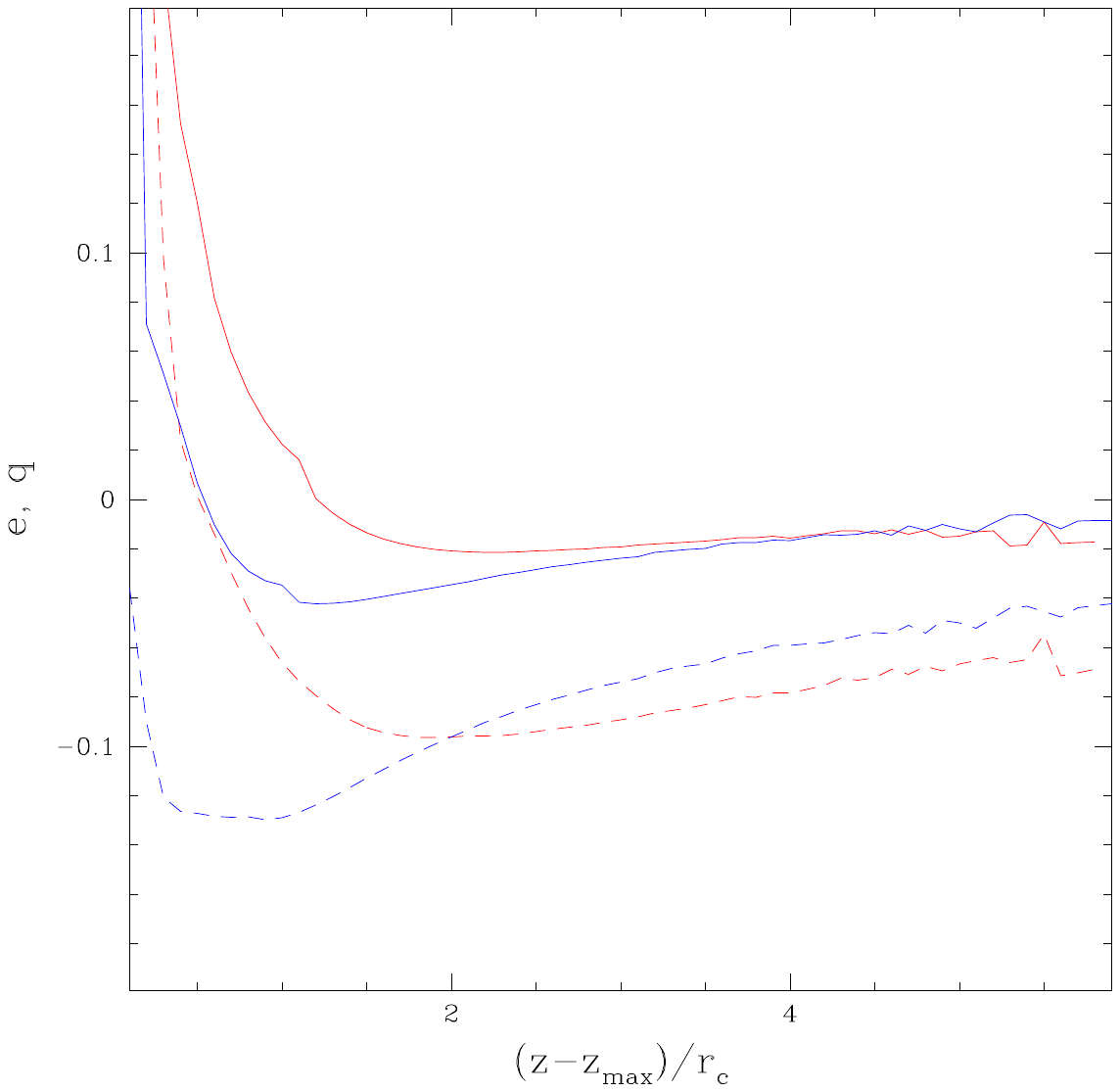}
\caption{Flattening (blue lines) and asymmetry (red lines) profiles for 
the isotropic (solid lines) and anisotropic (dashed lines) reference models.}
\label{mod2_fig}
\end{figure}

\begin{figure*}
\includegraphics[trim={0 5.5cm 0cm 2cm},clip,width=\textwidth]{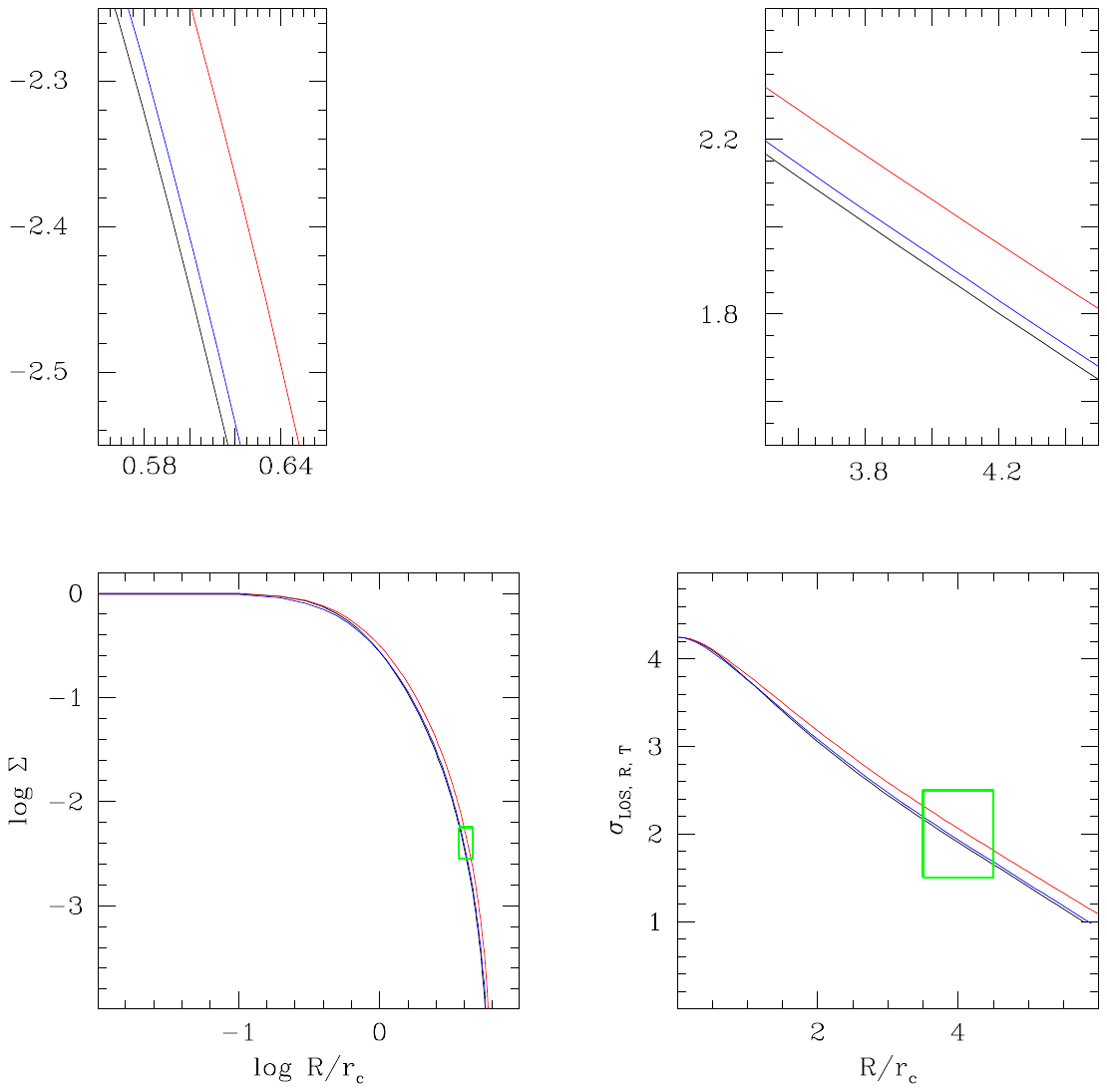}
\caption{Projected density (lower left panel) and velocity dispersion (lower right panel) profiles of the isotropic reference model, along the positive (black curves) and negative (red curves) branches of the $Z$ axis, and along the $X$ axis (blue curves). We recall that the positive $Z$ axis is parallel to and oriented as the external field (see text).
The upper panels show zooms of the green regions marked in the corresponding bottom panels.}
\label{mod3_fig}
\end{figure*}

\begin{figure*}
\includegraphics[trim={0 5.5cm 0cm 2cm},clip,width=\textwidth]{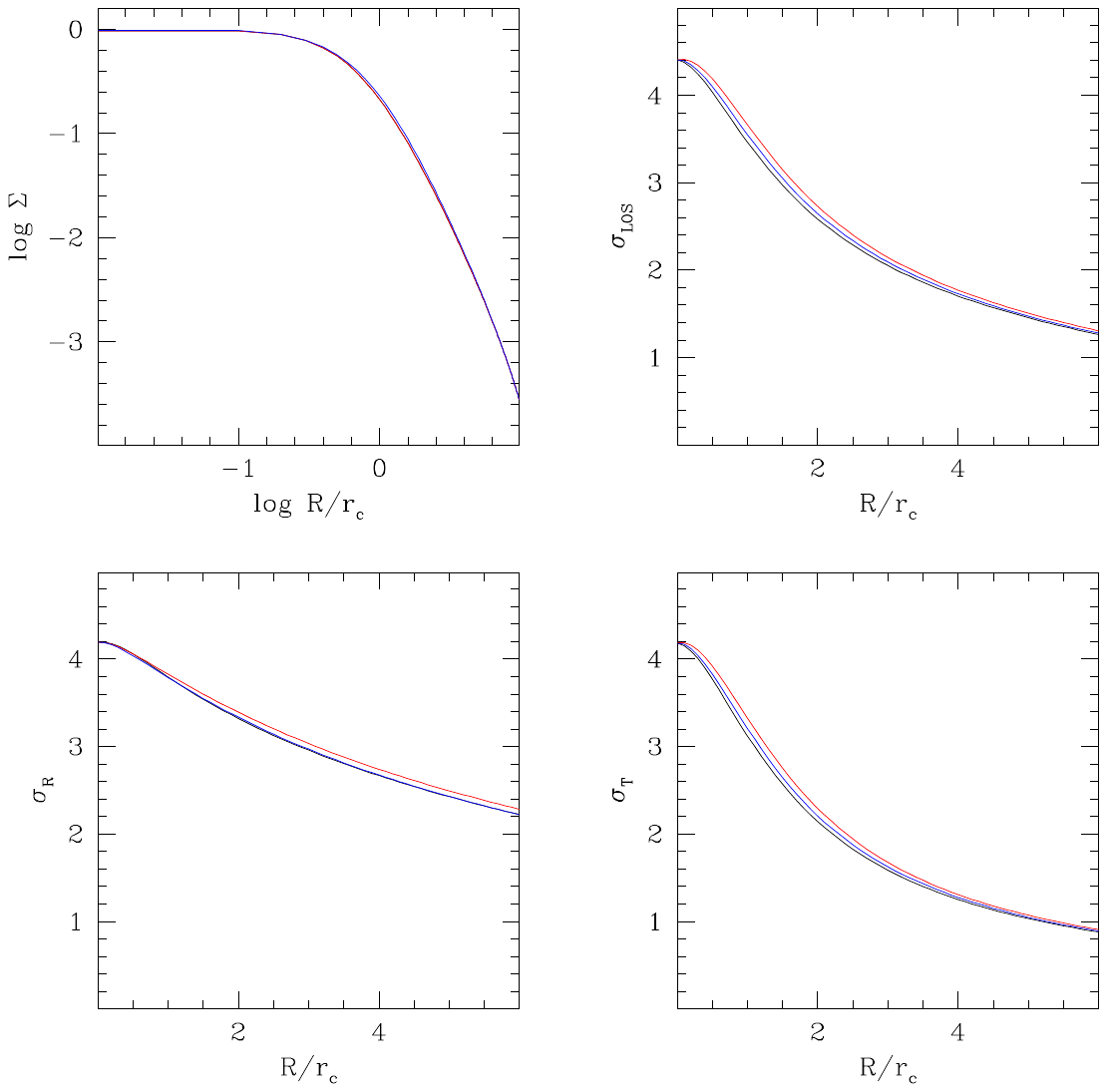}
\caption{Same as the lower panels of Fig.\ \ref{mod3_fig}, but for the anisotropic reference model.
Here, given that the model is anisotropic, we show separately the line-of-sight, and plane-of-the-sky radial and tangential velocity dispersion profiles.}
\label{mod3a_fig}
\end{figure*}

\begin{figure*}
\includegraphics[trim={0 5cm 7cm 2cm},clip,width=0.8\textwidth]{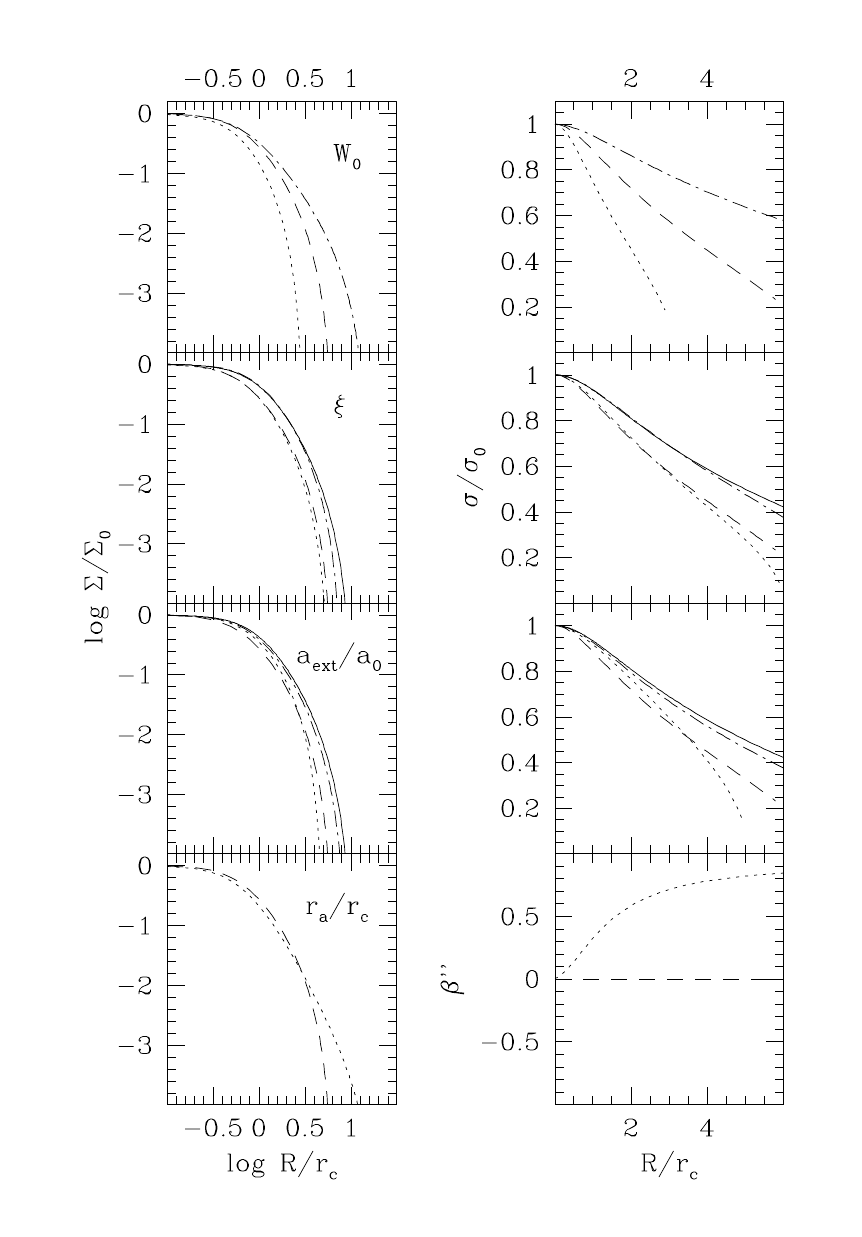}
\caption{Projected density (left panels) and velocity dispersion (right panels) profiles
of QUMOND models which are obtained from the isotropic reference model by varying  one parameter at a time.
From top to bottom: $W_{0}=3,~5,~7$, $\xi=0.1,~1,~10$, $\tilde{a}=0.1,~1,~10$ (marked in each panel with dotted, dashed and dot-dashed lines, respectively) and $\tilde{r}_{a}=1.3,~\infty$ (marked with  dotted and dashed lines, respectively). 
In the panels relative to the parameters $\xi$ and $\tilde{a}$, the Newtonian model with the same value of $W_{0}$ is also plotted 
with solid lines.}
\label{mod4_fig}
\end{figure*}

\begin{figure*}
\includegraphics[trim={0 6cm 0cm 13cm},clip,width=\textwidth]{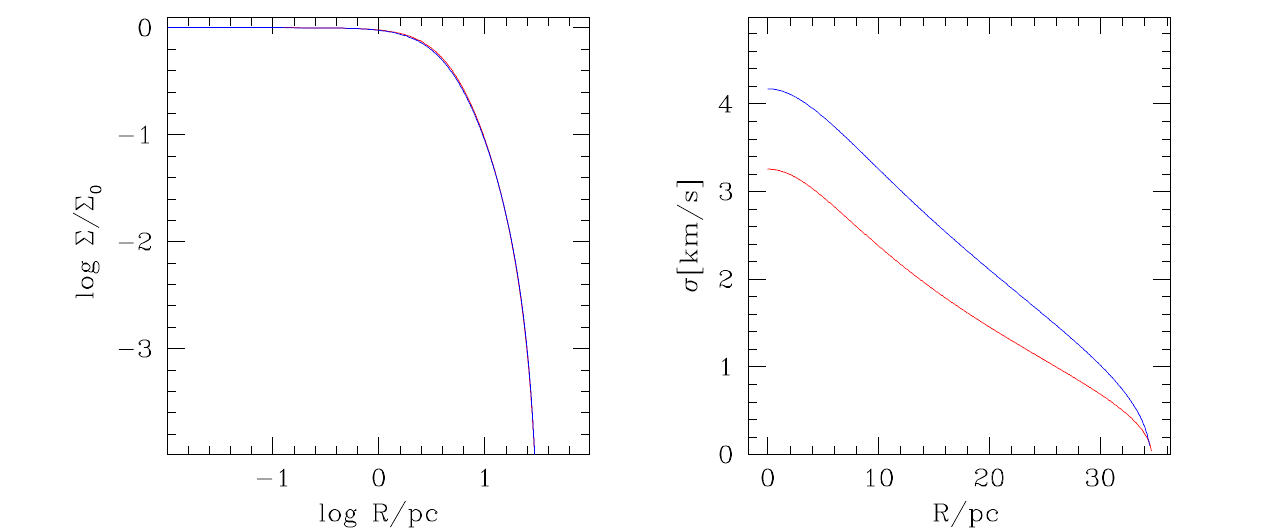}
\caption{Projected density (left panel) and velocity dispersion (right panel) profiles of the QUMOND isotropic reference model (blue lines) and 
of the Newtonian model with the same projected density profile (red lines).}
\label{mod5_fig}
\end{figure*}

To illustrate the characteristics of our models, we start with two reference 
sets of parameters corresponding to an isotropic and a maximally radially anisotropic model.
Both reference models have a mass of $10^{5} M_{\odot}$, a central dimensionless potential $W_{0}=5$, 
a MOND parameter $\xi=1$ and are immersed in a uniform external field with magnitude $\tilde{a}=1$.
So, the only varying parameter is $\tilde{r}_{a}$ which is obviously set to $\infty$ in the 
isotropic case and to $\tilde{r}_{a,min}=1.3$ in the anisotropic case. 
This value corresponds to a value of the Fridman-Poliachenko index 
$\zeta$ lower than $1.7$, which is  
the maximum value for which an anisotropic system remains stable against bar instability \citep{nipoti2011}. 
We recall that $\zeta$ is a {\em global} measure of anisotropy, defined as the the ratio of kinetic energy in radial and tangential motions:
\begin{equation}
\zeta=\frac{2T_{r}}{T_{t}}=\frac{2\int_{0}^{\pi}\int_{0}^{\infty}\tilde{r}^2 \sin\theta \tilde{\rho} \sigma_{r}^{2} d\tilde{r} d\theta}{\int_{0}^{\pi}\int_{0}^{\infty}\tilde{r}^2 \sin\theta \tilde{\rho} \sigma_{t}^{2} d\tilde{r} d\theta}
\label{pol_eq}
\end{equation}
\citep{fridman1984}.
In the following, we will quantify the degree of {\em local} anisotropy using the parameter
\begin{equation*}
\beta=1-\frac{\sigma_{t}^{2}}{2\sigma_{r}^{2}}
\label{beta_eq} 
\end{equation*}
for the three-dimensional structure of the system, and, when dealing with projected quantities, its analogue 
\begin{equation*}
\beta''=1-\left(\frac{\sigma_{T}}{\sigma_{R}}\right)^{2},
\label{betas_eq} 
\end{equation*}
where $\sigma_{R}$ and $\sigma_{T}$ are, respectively, the radial and tangential components in the plane of the sky of the velocity dispersion integrated along the line of sight \citep{sollima2015}.

In the top panels of Fig.s \ref{mod1_fig} and \ref{mod1a_fig} the maps of density, 
potential, velocity dispersion and anisotropy in the meridional plane $(x,z)$
(with the positive $z$ axis with the same direction and orientation as the external field) 
are shown, for the isotropic and anisotropic models, respectively. It can be seen that both models are almost spherical.
A zoom of the above profiles along the $x$ and the $z$ axes inside the core is shown in the bottom panels of the same figures.
Here the small ($z_{peak}\sim0.1~r_{c}$) shift of the peak density toward the direction of the external field is noticeable.
The profiles of other quantities along the $x$ and $z$ axes are almost indistinguishable. 
As expected, in the anisotropic model the $\beta$ parameter progressively increases toward the outer region of the system.

Fig. \ref{mod2_fig} shows the flattening ($q$) and asymmetry ($e$) profiles, defined as 
\begin{eqnarray*}
q&=&1-x/z_{+},\nonumber\\
e&=&1-z_{-}/z_{+},\nonumber
\label{flat_eq}
\end{eqnarray*}
where $x,~z_{-}$ and $z_{+}$ are the distances from the density peak of a given density level along the $x$ axis, 
and along the negative and positive branches of the $z$ axis, respectively.
It can be noted that there is an inversion of the trends of these quantities moving from the centre to the outer part of the system.
In particular, in the very central region (at $r<0.5~r_{c}$) the system iss elongated in the direction of the external field, 
but outside this region the trend inverts reaching very small ($e,q<0.1$) flattening and asymmetry in the opposite direction.
The same trend is magnified in the anisotropic model, never reaching significant levels of flattening and anisotropy. On the basis of the above evidence of small deviations from spherical symmetry, we can consider the angular momentum as a quasi-conserved quantity and safely adopt eq. \ref{df_eq} also for QUMOND models (see Sect.\ \ref{mod_descr_sec}).

The deviations from central symmetry are even smaller when considering projected quantities.
In Fig.s \ref{mod3_fig} and \ref{mod3a_fig} the projected density and the three components of the velocity dispersion are shown along the $X,~Z_{-}$ and $Z_{+}$ directions.
Here, $X,~Z_{-}$ and $Z_{+}$ are the equivalent of the 3D distances $x,~z_{-}$ and $z_{+}$, but projected into the plane of the sky assuming a line of sight orthogonal to $z$ to maximize the flattening and the asymmetry. 
Note that all profiles are extremely similar with differences of the order of $0.1$\,dex in the 
logarithmic density and $<0.1$\,km/s in the velocity dispersion for both isotropic and anisotropic models.
In this case, the model appears slightly elongated in the direction of the external field.

We plot in Fig. \ref{mod4_fig} the projected density and velocity dispersion of the isotropic reference model by changing one parameter at time.
Here it can be visualized that $W_{0}$ mainly affects the model concentration (as in all canonical Newtonian models; see \citealt{king1966,gunn1979}) with models with high $W_{0}$ asymptotically approaching the isothermal sphere.
The parameter $\xi$ is instead an indicator of the internal acceleration field, 
and therefore determines the contribution of internal gravity to keep the system in the MOND regime.
Indeed, models with large values of $\xi$ progressively approach their Newtonian equivalent.
The same occurs for the $\tilde{a}$ parameter for the external acceleration: the stronger the external field, the closer the
system to the Newtonian model.
Finally, the parameter $\tilde{r}_{a}$ determines the degree of anisotropy, with lower values of 
$\tilde{r}_{a}$ corresponding to elongation of the velocity ellipsoids occurring at smaller distance from the centre. 

A different experiment is shown in Fig. \ref{mod5_fig}. Here the projected density profile of the QUMOND isotropic reference model is fit with a Newtonian model, and the Newtonian and MOND projected velocity dispersion profiles are compared.
It is apparent that the QUMOND model predicts a larger velocity dispersion across the entire extent of the system.
As already discussed in Sect.\ \ref{intro_sec}, this is a consequence of the stronger gravitational field predicted by 
MOND in regimes of low accelerations (eq.\ \ref{g_eq}). 
In MOND all cluster stars able to cross the low-acceleration region need more 
kinetic energy with respect to the classical Newtonian gravitation law. 
Any region of the cluster contains a fraction of these stars, so 
the velocity dispersion is inflated at all radii.

\section{Observational data}
\label{obs_sec}

\subsection{GC data}
\label{gc_sec}

\begin{figure}
\includegraphics[trim={0 4cm 0cm 4cm},clip,width=\columnwidth]{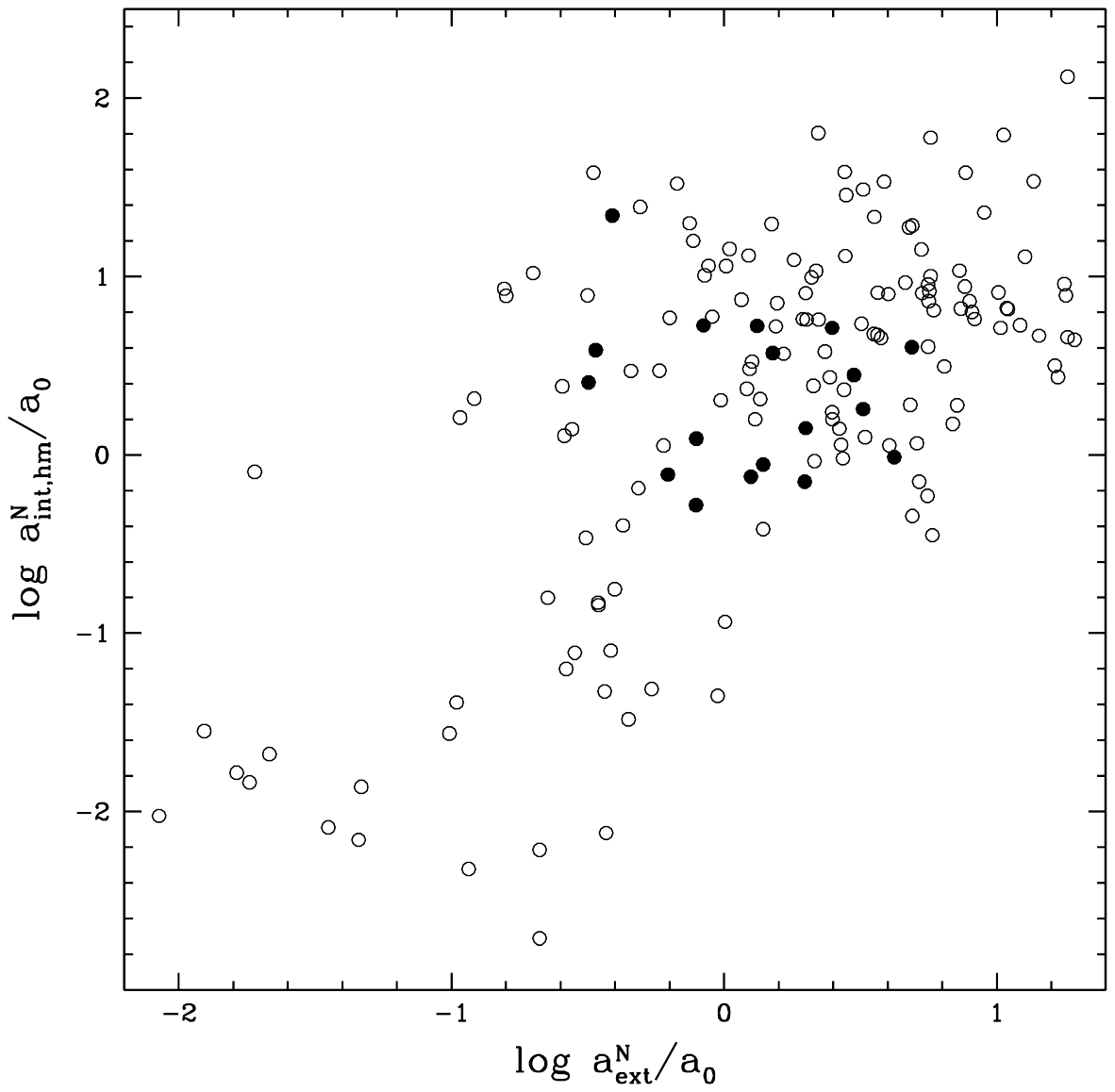}
\caption{Location of the 160 GCs of the \citet{baumgardt2018} database (open dots) in the $\log\,(a^{\rm N}_{int,hm}/a_{0})$ vs.\ $\log\,(a^{\rm N}_{ext}/a_{0})$ plane. 
The 18 GCs analysed in this paper are marked by full dots.}
\label{gc_fig}
\end{figure}

Among the 160 Galactic GCs, only a small subsample can be useful for our analysis.
 
The selection has been made on the basis of three different criteria: {\it i)} the absence of significant rotation, 
{\it ii)} the availability of accurate MFs sampled down to the least massive stars ($m\sim 0.1 M_{\odot}$), close to the hydrogen burning limit, 
{\it iii)} the availability of a significant number 
of accurate kinematic data (proper motions and line-of-sight velocities).

The first criterion is based on the fact that our models lack a treatment of internal rotation. 
Note that in some GCs rotation is significant, and the rotation and pressure supports are comparable \citep{sollima2019}.
So, we first exclude all those GCs which are classified as "significant" or "uncertain" rotators in any of the 
compilations of \citet{sollima2019} and \citet{vasiliev2021}.  

The second criterion is based on the approach we will adopt in Sect. \ref{tech_sec} to compute the consistency of our best-fit models with observations.
Indeed, we need to compare our dynamical $M/L$ with those obtained from the comparison of
the stellar population synthetically derived from stellar models. 
A fundamental input of these models is the MF which needs to be well sampled across the entire extent covered by each GC,
incuding the very low-mass stars contributing to the cluster mass more than to the light.   
We adopted the MF measurements by \citet{baumgardt2023} and restricted our sample to those GCs with a lower mass limit $m<0.26 M_{\odot}$. 

The third criterion is based on the statistical robustness of the derived $M/L$ and of the constraint on the anisotropy parameters.
We limit our analysis to GCs with at least 50 radial velocities and 50 proper motions with the required accuracy (see Sect.\ \ref{data_sec}). 

After the application of the above criteria, we selected a sample of 18 GCs.
They are listed in Table \ref{table1_tab}. They span a heliocentric distance range $1.8<R_{\odot}/\kpc<18.5$ and a Galactocentric distance range $2.5<R_{GC}/\kpc<18.5$.

\begin{table}
	\centering
	\caption{Properties of the Newtonian and QUMOND best-fit models. Column 1: name of the GC. Column 2: minimum $V$-band mass-to-light ratio (see Sect.\ \ref{mlmin_sec}). Columns 3 and 4: $V$-band mass-to-light ratio and probability (see Sect.\ \ref{mlgc_sec}) of the best Newtonian model. Columns 5 and 6: $V$-band mass-to-light ratio and probability of the best QUMOND model. Mass-to-light ratios are in solar units.}
	\label{table1_tab}
	\begin{tabular}{lccccr} 
		\hline
    & & \multicolumn{2}{c}{Newtonian} & \multicolumn{2}{c}{QUMOND}\\		
NGC & $(M/L_V)_{min}$ & $M/L_V$ & $P$ & $M/L_V$ & $P$\\
\hline
288  & 0.988 & 2.931 & 1.000 & 1.400 & 0.997\\
1261 & 0.811 & 1.896 & 1.000 & 1.097 & 0.965\\
1851 & 0.887 & 2.132 & 1.000 & 2.069 & 1.000\\
4590 & 1.187 & 2.960 & 1.000 & 1.883 & 0.984\\
4833 & 0.872 & 1.348 & 1.000 & 1.069 & 0.844\\
5024 & 1.232 & 2.028 & 1.000 & 0.927 & 0.005\\
5897 & 1.329 & 2.393 & 1.000 & 1.339 & 0.591\\
6101 & 1.329 & 2.568 & 1.000 & 1.386 & 0.640\\
6121 & 1.059 & 1.895 & 1.000 & 1.549 & 0.977\\
6171 & 0.953 & 2.078 & 1.000 & 1.518 & 1.000\\
6254 & 0.900 & 1.749 & 1.000 & 1.485 & 0.983\\
6352 & 0.943 & 2.067 & 0.999 & 3.198 & 1.000\\
6362 & 0.868 & 1.965 & 1.000 & 1.274 & 1.000\\
6366 & 0.720 & 1.564 & 1.000 & 1.077 & 0.994\\
6496 & 1.151 & 1.643 & 0.981 & 1.270 & 0.754\\
6723 & 0.927 & 2.232 & 1.000 & 1.697 & 1.000\\
6779 & 0.844 & 3.147 & 1.000 & 3.063 & 1.000\\
6838 & 0.725 & 1.207 & 1.000 & 1.191 & 1.000\\
\hline
	\end{tabular}
\end{table}

In Fig.\ \ref{gc_fig}, the 160 GCs of the \citet{baumgardt2018} sample  are plotted in the $\log\,(a^{\rm N}_{int,hm}/a_{0})$ vs.\ $\log\, a^{\rm N}_{ext}/a_{0}$ plane.
Here, 
\begin{equation*}
a^{\rm N}_{int,hm}=\frac{GM}{2~r_{h}^{2}}
\label{gint_eq}
\end{equation*}
is the Newtonian internal acceleration magnitude of the cluster measured at the half-mass radius $r_h$ and 
$a^{\rm N}_{ext}$ is the external acceleration magnitude calculated as described in Appendix \ref{sec:app_ext} (eq.\ \ref{atil_eq_app}).
Clusters distribute in this plane along a diagonal, because of the well-known Galactocentric distance-size relation \citep{vandenbergh1991}. 
It is apparent that all GCs with both $a^{\rm N}_{int,hm}<a_0$ and $a^{\rm N}_{ext}<a_0$ are outside the region where all the selection criteria are satisfied.
Indeed, they are too far to have accurate proper motions and a properly sampled MF.
Instead, those matching all the criteria occupy a region shared by many GCs between $0.32<\tilde{a}<4.88$.

\subsection{Surface density profiles}
\label{data_surf_den}

We account for the structural properties of the observed GCs, considering circularized surface density profiles.  
In particular, we adopt the surface density profiles of \citet{miocchi2013} when available and those of \citet{trager1995} otherwise.
For six GCs (NGC4833, NGC6101, NGC6352, NGC6362, NGC6496 and NGC6838) 
we calculated profiles using the ACS HST catalogs of \citet{anderson2008} for the cluster cores and those of \citet{stetson2019} for the outskirts.
For this purpose, we converted F606W HST magnitudes into Johnson ones using the transformations of \citet{sirianni2005}, and selected stars along the main sequence in the common magnitude interval $12<V<19$ where the photometric 
completeness is expected to be $>90\%$.
The surface density has been calculated by counting stars in circular annuli and dividing by the correspondent area.

\subsection{Proper motions and radial velocities}
\label{data_sec}

To account for the kinematic properties of the observed GCs, we rely mainly the proper motions provided by the 3rd data release of the Gaia survey \citep{gaia2021}
and the database of radial velocities collected by \citet{baumgardt2018} using a compilation of 
high-resolution spectroscopic data properly aligned.
We cross-matched the two data sets in order to obtain a single catalog per cluster containing all the three components 
of the velocities.

From this catalog we want to extract a selection of sufficiently accurate kinematic measurements for a subsample of bona-fide cluster members. 
For this purpose, we find it convenient to use as reference a Newtonian dynamical model of the cluster.  
We thus fit the surface density profiles (Sect.\ \ref{data_surf_den}) of each cluster with an isotropic ($r_{a}/r_{c}=\infty$) Newtonian \citet{gunn1979} model providing a 
normalized velocity dispersion ($\tilde{\sigma}_{v,i}^{2}$ and $\tilde{\sigma}_{\mu,i}^{2}$ for radial velocities and proper motions, respectively) 
at the projected radius of each star $R_{i}$.

To obtain the corresponding velocity dispersions in physical units $\sigma_{v,i}\equiv\sigma_{v}(R_i)$ and $\sigma_{\mu,i}\equiv\sigma_{\mu}(R_i)$, we need
two scaling factors ($\sigma_{v,0}$ and $\sigma_{\mu,0}$) 
such that 
$\sigma_{v,i}=\sigma_{v,0}\tilde{\sigma}_{v,i}$ and $\sigma_{\mu,i}=\sigma_{\mu,0}\tilde{\sigma}_{\mu,i}$. 
The relation between the two scaling factors is
\begin{equation}
\sigma_{v,0}=4.74\left(\frac{R_{\odot}}{\kpc}\right)\left(\frac{\sigma_{\mu,0}}{\mas/\yr}\right)\kms,
\label{conv_eq}
\end{equation}
where $R_{\odot}$ is the cluster distance. Throughout this work, we adopt the distances from \citet{baumgardt2021}, 
so in the following  $\sigma_{v,0}$ is left as a free parameter of the model, while $\sigma_{\mu,0}$ is obtained from $\sigma_{v,0}$ using eq.\ \ref{conv_eq}. 

The best-fitting isotropic Newtonian model is found by maximizing the likelihood 
\begin{equation}
\ln L_\Sigma=-\frac{1}{2}\sum_{j=1}^{M}\biggl(\frac{\log\Sigma_{{\rm obs},j} - \log\widetilde{\Sigma}(R_j) - \log \Sigma_0}{\delta\log\Sigma_{{\rm obs},j}}\biggr)^2, 
\label{eq:lnlsigma}
\end{equation}
where $\Sigma_{{\rm obs},j}$ is the observed surface density at radius $R_j$, $\delta\log\Sigma_{{\rm obs},j}$ is the uncertainty on $\log\Sigma_{{\rm obs},j}$, $\widetilde{\Sigma}(R)$ is the normalized model's surface density at projected radius $R$ and $\Sigma_0$ is the central surface density of the model, which is left as a free parameter.  

Then, we fit iteratively the mean cluster velocity components 
($\langle \mu_{\alpha}^{*}\rangle,~\langle\mu_{\delta}\rangle$ and $\langle v\rangle$) starting from the initial guesses of \citet{vasiliev2021}, 
together with the scaling factor $\sigma_{v,0}$, selecting those providing the maximum likelihood defined as 
\begin{eqnarray}
\ln L_{kin}&=&\ln L_{v}+\ln L_{\mu},\nonumber\\
\ln L_{v}&=&\sum_{i=1}^{N} \ln L_{v,i},\nonumber\\
\ln L_{v,i}&=&-\frac{1}{2}\left[\delta_{v,i}^{2}+\ln s_{v,i}^{2}+\ln(2\pi)\right],\nonumber\\
\ln L_{\mu}&=&\sum_{i=1}^{N} \ln L_{\mu,i},\nonumber\\
\ln L_{\mu,i}&=&-\frac{1}{2}[\delta X_{i}^{2}+\delta Y_{i}^{2}-2\tilde{\varrho}_{i}\delta X_{i}\delta Y_{i}+\ln(1-\tilde{\varrho}_{i}^{2})+\nonumber\\
&&\ln(s_{\mu X,i}^{2}s_{\mu Y,i}^{2})]-\ln(2\pi),
\label{like_eq}
\end{eqnarray}
where
\begin{eqnarray*}
\delta X_{i}^{2}&=&\frac{(\mu_{\alpha,i}^{*}-\langle\mu_{\alpha}^{*}\rangle)^{2}}{(1-\tilde{\varrho}_{i}^{2})s_{\mu X,i}^{2}},\nonumber\\
\delta Y_{i}^{2}&=&\frac{(\mu_{\delta,i}-\langle\mu_{\delta}\rangle)^{2}}{(1-\tilde{\varrho}_{i}^{2})s_{\mu Y,i}^{2}},\nonumber\\
s_{\mu X,i}^{2}&=&\epsilon_{\mu \alpha,i}^{2}+\sigma_{\mu,i}^{2},\nonumber\\
s_{\mu Y,i}^{2}&=&\epsilon_{\mu \delta,i}^{2}+\sigma_{\mu,i}^{2},\nonumber\\
\tilde{\varrho}_{i}&=&\frac{\varrho_{\alpha\delta,i}~\epsilon_{\mu \alpha,i}\epsilon_{\mu \delta,i}}{s_{\mu X,i}s_{\mu Y,i}},\nonumber\\
\delta v_{i}^{2}&=&\frac{(v_{i}-\langle v\rangle)^{2}}{s_{v,i}^{2}},\nonumber\\
s_{v,i}^{2}&=&\epsilon_{v,i}^{2}+\sigma_{v,i}^2).\nonumber
\label{muc_eq}
\end{eqnarray*}
Here $N$ is the number of bona-fide cluster members at the current iteration, 
$\mu_{\alpha,i}^{*},~\mu_{\delta,i}$ and $v_{i}$ are the proper motions and radial velocity of the $i$th star,
$\epsilon_{\mu \alpha,i},~\epsilon_{\mu \delta,i}$ and $\epsilon_{v,i}$ are their respective uncertainties, $\varrho_{\alpha\delta,i}$ is the correlation coefficient between $\mu_{\alpha,i}^{*}$ and $\mu_{\delta,i}$.
We adopted Powell's gradient descent algorithm \citep{powell1964} to find the maximum 
likelihood in the considered 4-dimensional parameter space.
At each iteration, we eliminated from the sample of bona-fide cluster members all those stars with velocity lying at more than 5$\sigma$ in the model's velocity distribution at their radius.
The algorithm converges after $\sim$10 iterations, providing the systemic motion of the cluster 
($\langle \mu_{\alpha}^{*}\rangle,~\langle\mu_{\delta}\rangle$ and $\langle v\rangle$) and the central scaling factor of radial velocity
($\sigma_{v,0}$).
This value has been converted, using eq.\ \ref{conv_eq} and 
the distances provided by \citet{baumgardt2021}, into proper motion scaling factor $\sigma_{\mu,0}$, which we adopted as upper limit in proper motion uncertainty ($\epsilon_{\mu,max}=\sigma_{\mu,0}$). The parallaxes of member stars have been also used to determine the systemic cluster parallax ($\langle p\rangle$) and its dispersion ($\sigma_{p}$).

All the Gaia proper motions of stars contained within the tidal radius of the best-fit \citet{gunn1979} model have been selected.
Among them, we selected for our final sample the stars {\it i)} lying along the characteristic sequences of the $G,~(G_{BP}-G_{RP})$ colour-magnitude diagram, 
{\it ii)} with a parallax contained within $5\sigma_{p}$ from the mean systemic cluster parallax, {\it iii)} with $\ln L_{\mu,i}>\ln L_{\mu,best}-5$ (where $L_{\mu,best}=\max_i L_{\mu,i}$), and {\it iv)} with $\min(\epsilon_{\mu \alpha,i},~\epsilon_{\mu \delta,i})<\epsilon_{\mu,max}$.  
Of course, from eq.\ \ref{conv_eq}, it is apparent that proper motions and their associated errors
are proportional to the cluster distance. So, the more distant is the target GC the larger will be
its proper motion uncertainty. Consequently, criterion {\it (iv)} greatly reduces the number of suitable proper motions for distant clusters.

In the next steps (Sect. \ref{tech_sec}) we use separately the sample of $N_v$ line-of-sight velocities 
of bona-fide members and the Gaia sample of proper motions for $N_{\mu}$ stars selected according to the above criteria. 

\section{Technique}
\label{tech_sec} 

In this Section, we describe how we derived for each cluster the dynamical $M/L_{V}$ predicted by the two different gravitation theories and its minimum value independently derived from stellar evolution models.

\subsection{Model fit and dynamical $M/L_{V}$}
\label{mlgc_sec}

\begin{figure*}
\includegraphics[trim={0 0cm 0cm 0cm},clip,width=\textwidth]{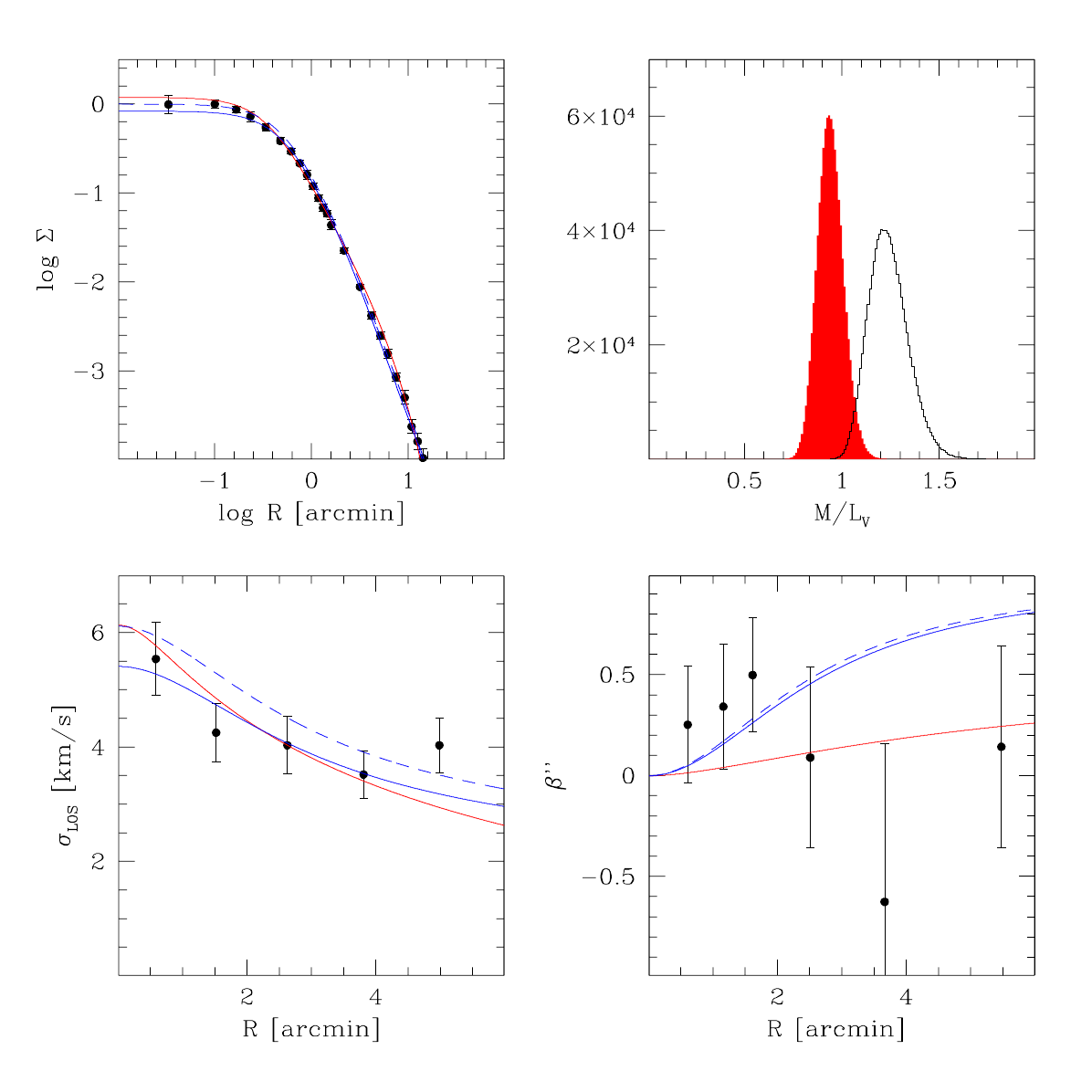}
\caption{Projected density (upper left panel), line-of-sight velocity dispersion (lower left panel) and 
projected anisotropy parameter (lower right panel) profiles of the best-fitting Newtonian (red solid curves) and QUMOND (blue solid curves) models of NGC 5024. The dashed blue curve indicates the best-fitting  QUMOND model assuming the minimum mass-to-light ratio $(M/L_V)_{min}=1.232$.
The black dots mark binned observational data for comparison, but the analysis has been conducted using unbinned data.
The probability distributions of $M/L_V$ for QUMOND models and of $(M/L_V)_{min}$ are shown in the upper right panel with red and empty histograms, respectively. Mass-to-light ratios are in solar units.}
\label{obsfit_fig}
\end{figure*}

Proper motions have been corrected for perspective
rotation using eq.\ 1 of \citet{gaia2018b} and eq.s 4 and 6 of \citet{vandeven2006}.
The celestial coordinates (RA, Dec) have been converted into projected distances
from  the  cluster centre using  equation  1  of  \citet{vandeven2006}
and adopting the centres of \citet{baumgardt2018}.  
The $\alpha$ and $\delta$ proper motions of each star have been converted into radial ($\mu_R$) and tangential ($\mu_T$) proper motions in the plane of the sky, relative to the cluster centre. 

The models have been projected along the line of sight, taking into account that the model symmetry axis $z$ (aligned with the external field, which points towards the Galactic centre) forms with the line of sight an angle
\begin{equation*}
i=\cos^{-1}\left(\frac{{\bf R_{GC}}\cdot{\bf R_{\odot}}}{\|{\bf R_{GC}}\|\|{\bf R_{\odot}}\|}\right),
\label{i_eq_app}
\end{equation*}
where ${\bf R_{GC}}$ and  ${\bf R_{\odot}}$ are the vectors connecting the cluster to the Galactic centre and to the Sun, respectively.
Given that the projections of the models in the plane of the sky deviate negligibly from circular symmetry (see Sect.\ \ref{mod_prop_sec}), for comparison with the data we computed for the model the angle averaged profiles of the following quantities integrated along the line of sight: the surface density $\Sigma(R)$, the line-of-sight velocity dispersion $\sigma_{\rm LOS}(R)$, and the radial $\sigma_{\mu R}(R)$ and tangential $\sigma_{\mu T}(R)$ components of the proper-motion dispersion, calculated using the cluster distance $R_{\odot}$ (here $R$ is the distance from the centre in the plane of the sky).

For both Newtonian and QUMOND models we defined a sequence of values of $M$ and $\tilde{r}_{a}^{-1}$ from 0 to $\tilde{r}_{a,min}^{-1}$ in steps of 0.1, where $\tilde{r}_{a,min}$ is such that $\zeta\simeq 1.7$ (see Sect.\ \ref{mod_prop_sec}).
As in Sect.\ \ref{data_sec}, the normalization factor of proper motions has been calculated separately and used only to calculate the contribution of anisotropy to the likelihood. 
Then, for each pair $(M,\tilde{r}_{a})$, using a Powell's gradient descent algorithm
\citep{powell1964}, we searched for the pair of values of parameters, ($W_{0},~r_{c}$) for Newtonian models and ($W_{0},~\xi$) for QUMOND ones, 
that maximize the following likelihood:
\begin{equation}
\ln L=\ln L_\Sigma+\ln L_{v}+\ln L_{\mu},
\label{like_rad_tan_eq}
\end{equation}
where 
$\ln L_\Sigma$ is defined in eq.\ \ref{eq:lnlsigma},
\begin{eqnarray}
\ln L_{v}&=&\sum_{i=1}^{N_v} \ln L_{v,i},\nonumber\\
\ln L_{\mu}&=&\sum_{i=1}^{N_{\mu}} \ln L_{\mu,i},\nonumber\\
\end{eqnarray}
and $\ln L_{v,i}$ and $\ln L_{\mu,i}$ are defined as in eq.\ \ref{like_eq}, but with 
\begin{eqnarray*}
\delta X_{i}^{2}&=&\frac{(\mu_{R,i}-\langle\mu_{R}\rangle)^{2}}{(1-\tilde{\varrho}_{i}^{2})s_{\mu X,i}^{2}},\nonumber\\
\delta Y_{i}^{2}&=&\frac{(\mu_{T,i}-\langle\mu_{T}\rangle)^{2}}{(1-\tilde{\varrho}_{i}^{2})s_{\mu Y,i}^{2}},\nonumber\\
s_{\mu X,i}^{2}&=&\epsilon_{\mu R,i}^{2}+\sigma_{\mu R}^{2}(R_{i}),\nonumber\\
s_{\mu Y,i}^{2}&=&\epsilon_{\mu T,i}^{2}+\sigma_{\mu T}^{2}(R_{i}),\nonumber\\
\tilde{\varrho}_{i}&=&\frac{\varrho_{RT,i}~\epsilon_{\mu R,i}\epsilon_{\mu T,i}}{s_{\mu X,i}s_{\mu Y,i}},\nonumber\\
\delta v_{i}^{2}&=&\frac{(v_{i}-\langle v\rangle)^{2}}{s_{v,i}^{2}},\nonumber\\
s_{v,i}^{2}&=&\epsilon_{v,i}^{2}+\sigma_{\rm LOS}^{2}(R_{i}),\nonumber
\label{muc_eq}
\end{eqnarray*}
where $\mu_{R,i}$ and $\mu_{T,i}$ are, respectively, the radial and tangential proper motions of the $i$th star,
$\epsilon_{\mu R,i}$ and $\epsilon_{\mu T,i}$ are the corresponding uncertainties, and $\varrho_{RT,i}$ is the correlation coefficient between $\mu_{R,i}$ and $\mu_{T,i}$.

By marginalizing over $\tilde{r}_{a}^{-1}$ we obtain the global likelihood for a given cluster mass $M$:  
\begin{equation*}
\mathcal{L}(M)=\int_{0}^{\tilde{r}_{a,min}^{-1}} L(M,\tilde{r}_a^{-1})~d\tilde{r}_{a}^{-1}.
\label{marg_eq}
\end{equation*}  

Masses have been then divided by the cluster luminosity, derived using the absolute $V$-band magnitudes by 
\citet{baumgardt2020} and the solar absolute $V$ magnitude $M_{V}=4.84$ \citep{prsa2016}, to obtain the 
corresponding distribution of $\mathcal{L}$ for $M/L_{V}$, which is then normalized and fitted with a Gaussian.
The best fit of the data of NGC5024 with Newtonian and QUMOND models are shown in Fig. \ref{obsfit_fig}, as an example. 
The dynamical $M/L_V$ of our sample of GC are reported in Table \ref{table1_tab}.

\subsection{Minimum $M/L_{V}$}
\label{mlmin_sec}

The goal of this paper is to compare the dynamical $M/L_{V}$ of our GCs sample with a dynamics-independent 
estimate, to test the validity of the Newtonian and QUMOND theories of gravitation.
In particular, it is important to estimate an observationally inferred minimum $M/L_{V}$ of the GC, independent of kinematics:
as pointed out in the Introduction, for a gravitational theory to be acceptable, the dynamical $M/L_{V}$ predicted by the theory must not be lower than this minimum value. Thus the estimate of the minimum $M/L_{V}$ is a fundamental piece of the present investigation. 

A viable option is provided by the $M/L_{V}$ predicted by stellar evolution models.
Indeed, each cluster star contributes to both mass and luminosity in a 
different way according to its initial mass and evolutionary stage.
As a first step, we choose a set of isochrones from the \citet{cassisi2000} database with 
suitable metal content $Z$ and age.
These isochrones use a solar mixture and extend from very low mass stars 
($m\sim 0.1 M_{\odot}$) to asymptotic 
giant branch stars and include mass loss occurring during cluster evolution.
The metal content has been derived using the metallicities $[Fe/H]$ from the \citet[][2010 edition]{harris1996}
catalog, accounting for the effect of $\alpha$-enhancement using the relation from \citet{salaris1993}
\begin{equation*}
\log Z=\log(0.02)+[[Fe/H]+\log(0.638~f_{\alpha}+0.362)]
\label{z_eq}
\end{equation*}
with
\[ f_{\alpha} = 
 \begin{cases}
  10^{0.28} & \quad \text{if } [Fe/H]<-0.8\\
  10^{-0.35~[Fe/H]} & \quad \text{if } [Fe/H]>-0.8.
 \end{cases}
\]

The age of each cluster has been derived by converting colors and magnitudes of 
isochrones of different ages into absolute magnitudes and dereddened colors using the distance of \citet{baumgardt2021}, 
the reddening of \citet[][2010 edition]{harris1996} and the extinction coefficients by \citet{cardelli1989}.
We choose the age providing the lowest $\chi^{2}$, calculated using the stars within 2 magnitudes from 
the turn-off point.

As stars of different masses contribute to the cluster mass and luminosity budget in a different way,
it is essential to know their relative fraction (the present-day MF).
We model the MF as a
single power-law with slope $\alpha_{MF}$, which has been shown to be a good approximation for many GCs \citep{ebrahimi2020}.
In particular, we adopt the MF measured by \citet{baumgardt2023}.

We assume that mass dependent depletion of stars has turned an initial \citet{kroupa2001} MF into the
observed MFs.
We model the the passive evolution of the initial population, using the relations of \citet{kruijssen2009}. According to these relations, stars above 8~$M_{\odot}$ evolve into neutron stars or into black holes depending on their original mass.
However, most of these stars are expected to quickly escape from the system because of the natal kick occurring at the end of their evolution \citep{drukier1996}.
As we want to estimate the minimum $M/L_V$, we exclude all stars with initial masses $m>8~M_{\odot}$. 
The stars with $m<8~M_{\odot}$ become white dwarfs and do not suffer from natal kicks.
The mass in main-sequence stars is computed by integrating the present-day MF between $0.1~M_{\odot}$ and the turn-off mass.
To this mass, we add the mass in white dwarfs, computed assuming that white dwarfs are lost at the same fraction as main-sequence stars of the same mass, and using the initial-final mass relation of \citet{kalirai2008}.  
The luminosities of all stars, derived from the best-fit isochrone, 
have been finally summed to provide $L_V$, and thus an estimate of $M/L_V$ which is independent of the cluster dynamics.

Note that this $M/L_V$ is a lower limit to the actual value, mainly because the mass in white dwarf is a lower limit. White dwarfs are being lost at a lower rate compared to main-sequence stars of the same mass, since they are more massive stars for a significant fraction of the time and, also when they turn into white dwarfs, they start from the centre, so it takes them a long time to drift towards the tidal radius. $N$-body simulations have shown that white dwarfs can contribute up to 70\% of the total mass in an evolved cluster \citep[see Fig. 11 of ][]{baumgardt2003}. Instead, in our estimates of the present-day cluster mass, in which this effect is neglected, the contribution of white dwarfs never exceeds 25\%.

For each cluster we repeat the above task $10^{3}$ times by adding to cluster distances 
and metallicities a random shift extracted by Gaussian distributions centred on the nominal value and with a standard deviation 
equal to the $1\sigma$ uncertainties quoted by \citet{baumgardt2021} for distances and a typical value of 0.1 dex for metallicities.
The distribution of the output $(M/L_V)_{min}$ has been assumed as representative of the 
probability distribution of $(M/L_V)_{min}$.

For each GC $10^{6}$ pairs (dynamical $M/L_V$ and $(M/L_V)_{min}$) have been extracted from the estimated distributions and the fraction of occurrences of dynamical $M/L_V>(M/L_V)_{min}$ has been assumed as the probability $P$ of compatibility between the data and the considered model.

\section{Results}
\label{res_sec} 

\begin{figure*}
\includegraphics[trim={1cm 5.5cm 0cm 10cm},clip,width=\textwidth]{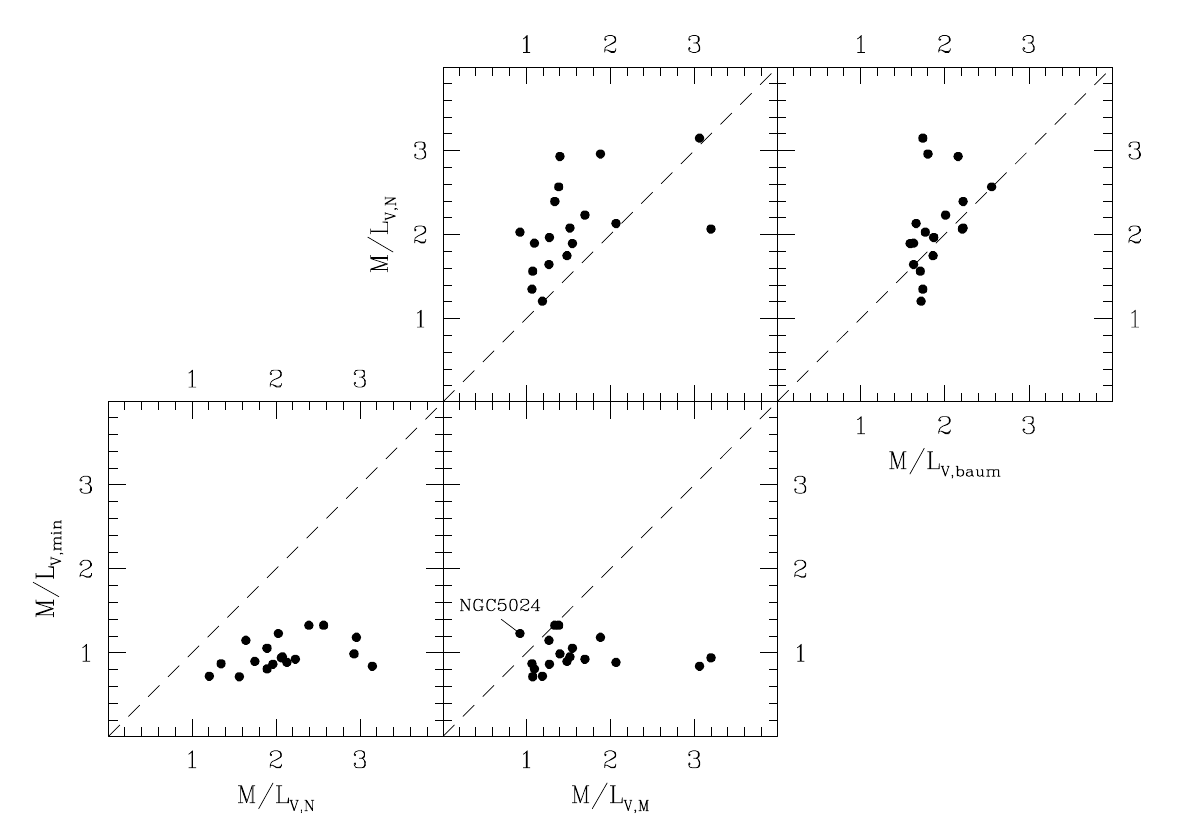}
\caption{Comparison among different estimates of $M/L_V$ (in solar units) for our sample of GCs. In the upper panels we compare the best-fitting Newtonian $M/L_V$ with that of QUMOND (left panel) and that of \citet{baumgardt2021} (right panel).
In the lower panels we compare the minimum $(M/L_V)_{min}$ with the best-fitting $M/L_V$ of Newtonian (left panel) and QUMOND (right panel) models.
The location of NGC5024 in the $(M/L_V)_{\rm M}$ vs.\ $(M/L_V)_{min}$ plane is shown.}
\label{ml_fig}
\end{figure*}

The probability of agreement $P$ for the 18 GCs of our sample is listed in Table \ref{table1_tab} for both Newtonian and QUMOND models.
None of the analysed clusters has been found with a dynamical $M/L_V$ 
significantly incompatible ($P<0.003$, corresponding to $\sim3\sigma$) with the predicted lower limit. 
For one of them (NGC5024) the QUMOND prediction lies at $2.8\sigma$ below such a lower limit ($P=0.005$).

In Fig. \ref{ml_fig} the dynamical $M/L_V$ estimated by Newtonian and QUMOND models are compared with $(M/L_V)_{min}$ and with the mass-to-light ratios measured by \citet{baumgardt2021}.
The $M/L_V$ of our Newtonian models are in good agreement with those of \citet{baumgardt2021}, with the exception of 3 GCs (NGC288, NGC4590 and NGC6779).

As expected, the Newtonian models are systematically more massive than QUMOND ones.
As already discussed in Sect. \ref{intro_sec}, this is a direct consequence of the increased acceleration in MOND models, which therefore require less mass to keep the cluster in equilibrium.
An exception is the cluster NGC6352, for which the Newtonian $M/L$ is lower than the QUMOND $M/L$: this can be explained by the fact that for this cluster the QUMOND best fit has a lower anisotropy than the Newtonian one.  

When comparing the dynamical $M/L_V$ with the minimum ones estimated from 
stellar evolution, it is apparent that the $M/L_V$ of Newtonian models are systematically higher than $(M/L_V)_{min}$, indicating a consistency between these models and independent observational constraints.
The $M/L_V$ of QUMOND models are on average closer to $(M/L_V)_{min}$. Remarkably, in QUMOND the cluster NGC5024 has a best-fitting dynamical $M/L_V$ lower than $(M/L_V)_{min}$, thus showing a deficiency of mass (though contained within the uncertainties).

\section{Conclusions}
\label{concl_sec}

In this paper we developed dynamical models of stellar systems within the framework of the 
quasi-linear modified Newtonian dynamics (QUMOND), which include radial anisotropy in the same fashion as their analogues in the standard Newtonian dynamics \citep{gunn1979}.
We compared them with the most updated set of observational 
kinematics of a sample of 18 GCs located in the Galactic halo in a range of Galactocentric distances $2.5<R_{GC}/\kpc<18.5$, a region characterized by external accelerations in the range
$0.32<a^{\rm N}_{ext}/a_{0}<4.88$ and derived their dynamical $M/L_V$.
We then tested the validity of both Newtonian and QUMOND theories by comparing these values with independent lower limits derived through the use of synthetic stellar evolution models.

As reported in Sect.\ \ref{res_sec}, none of the analysed clusters has a dynamical $M/L_V$ 
formally incompatible ($>3\sigma$) with the minimum $(M/L_V)_{min}$ prediction of 
stellar evolution models,
although one of them (NGC5024) reaches a disagreement with the QUMOND 
prediction at $2.8\sigma$.

Considering that there is still room to improve the accuracy of the estimated $M/L_V$ and their corresponding 
lower limits, it is possible that the incompatibility of this cluster (or others not sampled by our survey)
could become significant when better data are available.

A limit of our QUMOND model is that we neglect the fact that the external field varies while a GC moves along its orbit. In particular, if the external field experienced by NGC5024 was stronger in the past, the kinematics of this object might
retain memory of when it was in a more Newtonian regime, which in principle could help explain the low dynamical $M/L_V$ found for our QUMOND model. However, this effect is expected to be unimportant because (i)
the timescale over which the GC reacts to any change of the external field is typically much shorter than the orbital timescale \citep[even for very eccentric orbits;][]{wu2013} and (ii)  NGC5024 should not experience strong variations of the external field because its orbit has a  relatively low eccentricity of $\approx 0.4$  \citep[][]{vasiliev2021}.

Binaries cannot solve the discrepancy found for NGC5024: the effect of such objects is to inflate the actual velocity dispersion because of the velocity oscillation of the primary component around the centre of mass that 
spuriously adds a spread to the actual velocity dispersion \citep[see e.g.][]{bradford2011}.
So, the net effect would be to further decrease the required mass (and consequently the $M/L_V$)
needed to fit observations.

Similar considerations hold for the tidal heating. Also in this case, the kinetic energy released by the Galactic tidal field to the cluster stars would result in an increase of their velocity dispersion, 
thus enhancing the tension between predictions and observations \citep{spitzer1973}.

An opposite effect is instead produced by mass segregation. Indeed, the radial velocities available for most GCs are those of the red giant stars, which are the most massive stars of the sample.
These stars tend to sink in the central region of the cluster because they tend to release kinetic energy to less massive stars.
Therefore, they populate preferentially inner orbits with a velocity dispersion which, for a given mass, 
is lower than that predicted by single-mass models. So, by neglecting this effect, the best fit of the 
velocity dispersion neglects the contribution of the dynamically hot low-mass 
stars, resulting in an underestimated mass and $M/L_V$ up to a factor of $\sim$2 \citep{sollima2017a}.
For this reason, even a formal disagreement could not falsify MOND theories until multi-mass models in this gravitational framework are developed.
The models used in this paper can be generalized to allow for the presence of a spectrum of mass, as described in 
\citet{gunn1979}. However, multi-mass QUMOND models of GCs would require some inputs from simulations with QUMOND collisional $N$-body codes, which, as far as we know, have not been developed so far (see \citealt{ciotti2004} for a discussion of two-body relaxation in MOND). 

Summarizing, although the present analysis is not able to provide firm conclusions on the validity of QUMOND,
we show that this approach can be valuable for this purpose in the future.
Indeed, while previous analyses \citep{ibata2011a,sollima2012} were limited to only 2 
GCs subject to a negligible external field, here we can include GCs populating the inner Galactic halo, enlarging the number of target clusters.
Note that many GCs in this external acceleration range have promising 
properties (such as relatively steep MFs and low mass), but could not be included in our sample because of the lack of a significant number 
of radial velocities and/or proper motions with the required accuracy. The next releases of Gaia foresee an 
improvement in the accuracy and depth of proper motions \citep{gaia2018a}, and surveys of radial velocities are continuously in progress.
This could further enlarge the number of available target GCs and decrease the width of the $M/L_V$ probability distributions, thus improving the efficiency in detecting tensions between models and observations.
On the theoretical side, the development of multi-mass QUMOND models will account for the 
effects of mass segregation, providing a more complete representation of real GCs.

\section*{Acknowledgements}

We warmly thank Michele Bellazzini for useful discussions and suggestions.
We are grateful to an anonymous referee for constructive comments that helped improve the paper.

\section*{Data Availability}

The data underlying this article will be shared on reasonable request to the authors.



\bibliographystyle{mnras}




\appendix

\section{Computation of the models}
\label{sec:app_mod}

In this appendix we describe in more details the computation of the models presented in Sect. \ref{mod_sec}.

Reminding the normalizations of parameters given in eq. \ref{norm_eq},
the QUMOND modified Poisson equation of eq. \ref{eq:psimond} can be written as
\begin{subequations}
\begin{eqnarray}
\tilde{\nabla}^{2} W_{\rm M}&=&-9\nu\tilde{\rho}+\nu'\tilde{\nabla} y\cdot\left(\tilde{\nabla} W_{\rm N}+\frac{\tilde{a}}{\xi}\right)
\label{mond_pois_eq_app}\\
 &=&-9 \tilde{\rho_{f}}, \label{mond_pois_eqb_app}
\end{eqnarray}
\label{mond_pois_eq_app}
\end{subequations}
where $\tilde{\nabla}^2\equiv r_c^2\nabla^2$, $\tilde{\nabla}\equiv r_c\nabla$ and
\begin{equation}
\tilde{\rho}_{f}=\nu\tilde{\rho}-\frac{\nu'}{9} \tilde{\nabla} y \cdot \left( \tilde{\nabla} W_{\rm N}+\frac{\tilde{a}}{\xi} \right).
\label{rhof_eq_app}
\end{equation}
As far as we use the simple interpolating function $\mu(x)=x/(1+x)$, $\nu(y)$ is given by eq.\ \ref{nu_eq} and 
\begin{equation}
\nu'(y)=-\frac{1}{y\sqrt{y^{2}+4y}},
\label{nuder_eq_app}
\end{equation}
where  
\begin{equation}
    y=\|\xi\tilde{\nabla} W_{\rm N}+\tilde{a}\|
\end{equation}
in terms of dimensionless quantities.

We define the Newtonian and MOND potentials and densities as combinations Legendre polynomials 
(eq.\ \ref{defphi_eq})
\begin{eqnarray*}
W&=&\sum_{k=0}^{N} \tilde{u}_{k}(r) P_{k}(\theta),\nonumber\\
\tilde{\rho}&=&\sum_{k=0}^{N} \tilde{g}_{k}(r) P_{k}(\theta),\nonumber
\label{defphi_eq_app}
\end{eqnarray*} 
where we normalized the functions $u_{k}$ and $g_{k}$ as 
\begin{eqnarray*}
\tilde{u}_{k}&=&-\frac{u_{k}}{\sigma_{K}^{2}}\nonumber\\
\tilde{g}_{k}&=&\frac{g_{k}}{\rho_{0}}\nonumber
\label{normug_app*}
\end{eqnarray*}
or, in expanded form,
\begin{eqnarray}
\tilde{u}_{0}&=&W_{0}-9 \left(\int_{0}^{\tilde{r}} \tilde{r} \tilde{g}_{0}~d\tilde{r}-\frac{1}{\tilde{r}}\int_{0}^{\tilde{r}} \tilde{r}^{2} \tilde{g}_{0}~d\tilde{r}\right),\\
\tilde{u}_{k}&=&\frac{9}{2k+1} \left(\tilde{r}^{k}\int_{\tilde{r}}^{\infty} \tilde{r}^{1-k} \tilde{g}_{k}~d\tilde{r}+\tilde{r}^{-1-k}\int_{0}^{\tilde{r}} \tilde{r}^{k+2} \tilde{g}_{k}~d\tilde{r}\right),\nonumber\\
\label{uk_eq_app}
\end{eqnarray}
and
\begin{equation}
\tilde{g}_{k}=\frac{2k+1}{2} \int_{0}^{\pi} \tilde{\rho}~P_{k}~\sin\theta~d\theta.
\label{gk_eq_app}
\end{equation}
Note that there are two sets of coefficients $u_k$ and $g_k$ for Newtonian and QUMOND,
with the same functional definitions, but calculated using $\rho$ of $\rho_{f}$, respectively.
For simplicity, in the following we omit the suffixes N and M for $u_{k}$ and $g_{k}$, keeping in mind that these coefficients are calculated for both Newtonian and MOND models.

At the first iteration, we choose $W_{{\rm M},0},~\xi$ and $\tilde{a}$, 
and assume $N=0$.
This implies 
\begin{equation}
\tilde{g}_{0}=\tilde{\rho}
\label{g0_eq_app}
\end{equation}
and then
\begin{equation}
\tilde{u}_{0}=W_{{\rm N},0}-9 \left(\int_{0}^{\tilde{r}} \tilde{r} \tilde{\rho}~d\tilde{r}-\frac{1}{\tilde{r}}\int_{0}^{\tilde{r}} \tilde{r}^{2} \tilde{\rho}~d\tilde{r}\right).
\label{u0_eq_app}
\end{equation}
The above model is spherical and can be easily integrated from the centre outward.
Note that our input is $W_{{\rm M},0}$ while the Newtonian potential at the centre is unknown.
To overcome to this problem, we run a pre-iteration with $W_{{\rm N},0}=W_{{\rm M},0}$ and construct a model 
starting from the inner boundary conditions

\begin{eqnarray}
\tilde{\rho}&=&1,\nonumber\\
y&=&|\tilde{a}|,\nonumber\\
\frac{dW_{\rm N}}{d\tilde{r}}&=&0.
\label{bound_eq_app}
\end{eqnarray}

After substitution of variables and integration, eq.s \ref{dv3d}, \ref{dv3d_sigr} and \ref{dv3d_sigt}  can be written as functions of $W$ and $\tilde{r}_{a}$ as
\begin{equation}
\tilde{\rho}=\frac{\sqrt{\pi}e^{W}~\erf(\sqrt{W})+\frac{\sqrt{\pi}e^{-\tilde{r}_a^{2}W}~\erfi(\tilde{r}_a\sqrt{W})}{\tilde{r}_a^{3}}-\frac{2\sqrt{W}(1+\tilde{r}_a^{2})}{\tilde{r}_a^{2}}}{(1+\tilde{r}_a^{2})\left[\sqrt{\pi}~e^{W_{0}}~\erf(\sqrt{W_{0}})-\frac{2}{3}(2W_{0}+3)\right]},
\label{dv3d_eq_app}
\end{equation}
\begin{equation}
    \sigma_{r}^{2}=\sigma_{K}^{2}\frac{j_{0}-j_{2}}{(j_{0}+j_{2}\tilde{r_{a}}^{2})}
\end{equation}
and
\begin{equation}
\sigma_{t}^{2}=\sigma_{K}^{2}\frac{2j_{0}+j_{2}\left[(5+2W(1+\tilde{r_{a}}^{2}))\tilde{r_{a}}^{2}+3\right]-2W^{\frac{5}{2}}(1+\tilde{r_{a}}^{2})}{(j_{0}+j_{2}\tilde{r_{a}}^{2})(1+\tilde{r_{a}}^{2})},
\end{equation}
where
\begin{eqnarray*}
j_{0}&=&\frac{3}{4}\sqrt{\pi}e^{W}\erf(\sqrt{W})-\sqrt{W}(W+3/2),\nonumber\\
j_{2}&=&\frac{\frac{3}{4}\sqrt{\pi}e^{-\tilde{r_{a}}W}\erfi(\tilde{r_{a}}\sqrt{W})+\tilde{r_{a}}\sqrt{W}(\tilde{r_{a}}^{2}W-3/2)}{\tilde{r_{a}}^{5}},\nonumber
\label{j_eq_app}
\end{eqnarray*}
and $\erf$ and $\erfi$ are the real and imaginary error functions, respectively.

At each radial step, we calculate the quantities $\rho$, $\tilde{g}_{0}$, and
$\tilde{u}_{0}$ using eq.s \ref{dv3d_eq_app}, \ref{g0_eq_app}, and \ref{u0_eq_app}, respectively.
The value of $W$ at the origin of the axes is adopted as $W_{N,0}$, and a new iteration is started using only the first two equations of \ref{bound_eq_app}, 
until the value of $W_{N,0}$ converges within 1\%.

Once the spherical zero-model is computed, it is used to compute $u_{k}$ and $g_{k}$
for the desired value of $N$ through eq.s \ref{uk_eq_app} and \ref{gk_eq_app}, respectively.
For convenience, we report below the expansion in Legendre polynomials to compute the terms
$y$ and $\tilde{\nabla} y\cdot\left(\tilde{\nabla} W_{\rm N}+\frac{\tilde{\bf a}}{\xi}\right)$ needed in
eq.\ \ref{rhof_eq_app}:
\begin{equation*}
y=\sqrt{\left(\xi\sum_{k=0}^{N}\frac{du_{k}}{d\tilde{r}}P_{k}+\tilde{a}~\cos\theta\right)^{2}+\left(\frac{\xi}{\tilde{r}}\sum_{k=0}^{N}\frac{dP_{k}}{d\theta}u_{k}+\tilde{a}~\sin\theta\right)^{2}}\nonumber\\
\label{y_eq_app}
\end{equation*}
and
\begin{equation*}
\begin{aligned}
\tilde{\nabla} y&\cdot\left(\xi\tilde{\nabla} W+{\bf \tilde{a}}\right)=\frac{1}{y}\left[
\left(\xi\sum_{k=0}^{N}\frac{du_{k}}{d\tilde{r}}P_{k}+\tilde{a}~\cos\theta\right)^{2}\sum_{k=0}^{N}\frac{d^{2}u_{k}}{d\tilde{r}^{2}}P_{k}+\right.\nonumber\\
\frac{1}{\tilde{r}}&\left(\frac{\xi}{\tilde{r}}\sum_{k=0}^{N}\frac{dP_{k}}{d\theta}u_{k}+\tilde{a}~\sin\theta\right)^{2}\left(\frac{1}{\tilde{r}}\sum_{k=0}^{N}\frac{d^{2}P_{k}}{d\theta^{2}}u_{k}-\sum_{k=0}^{N}\frac{du_{k}}{d\tilde{r}}P_{k}\right)+\nonumber\\
\frac{2}{\tilde{r}}&\left.\frac{du_{k}}{d\tilde{r}}\frac{dP_{k}}{d\theta}\left(\xi\sum_{k=0}^{N}\frac{du_{k}}{d\tilde{r}}P_{k}+\tilde{a}~\cos\theta\right)\left(\frac{\xi}{\tilde{r}}\sum_{k=0}^{N}\frac{dP_{k}}{d\theta}u_{k}+\tilde{a}~\sin\theta\right)
\right].\nonumber\\
\end{aligned}
\label{prody_eq_app}
\end{equation*} 

As already discussed in Sect. \ref{mod_sec},
because of the presence of the external field, the isodensity surfaces are asymmetric
and elongated along the direction of the external field.
So, at each iteration, the density profiles are shifted to match the origin of axes with the centre of the system.
The new density map is used as input to compute the updated values of $u_{k}$ and $g_{k}$. The density profiles of different steps are then compared and a new iteration is started if the average variation exceeds 0.1\% of the central density.

\section{Computation of the external field}
\label{sec:app_ext}

The properties  of the external field for each GC depend on the adopted
Galactic model. For simplicity, in this work we adopt a spherical Galactic model.
This choice has some advantages. First, while in general
\begin{equation}
\nu(\|{\nabla\phi_{\rm N}}/a_{0}\|)\nabla\phi_{\rm N}=\nabla \phi_{\rm M}+\nabla\times {\bf A},
\label{extfield_eq_app}
\end{equation}  
where $\nabla\times{\bf A}$ is some unknown solenoidal field, in spherical symmetry the term $\nabla\times {\bf A}$ 
vanishes and eq.\ \ref{g_eq} can be properly used.
Second, in spherical symmetry the external field points everywhere towards the Galactic centre and its strength depends only on the Galactocentric distance.
We assume that the QUMOND acceleration must reproduce the rotation velocity of the Galactic disk \citep[$v_{rot}=229$ km/s;][]{eilers2019}.
So, from eq.\ \ref{extfield_eq_app}, assuming $\nabla\times {\bf A}=0$, we have
\begin{equation}
\nu(\|{\nabla\phi^{\rm MW}_{\rm N}}\|/a_{0}) \|{\nabla\phi^{\rm MW}_{\rm N}}\|=\frac{v_{rot}^{2}}{R_{GC}}
\label{atil_eq_app},
\end{equation}
where $\phi^{\rm MW}_{\rm N}(R_{GC})$ is the Newtonian potential generated by a density distribution with QUMOND potential  $\phi^{\rm MW}_{\rm M}(R_{GC})$ such that $\|\nabla\phi^{\rm MW}_{\rm M}\|={v_{rot}^{2}}/{R_{GC}}$.
Eq.\ \ref{atil_eq_app} can be solved numerically, providing for any given Galactocentric distance $R_{GC}$ the associated value of $\tilde{a}=a_{ext}^{\rm N}/a_0=\|\nabla\phi^{\rm MW}_{\rm N}\|/a_0$.


\bsp	
\label{lastpage}
\end{document}